\documentclass[journal=jpcafh]{achemso}
\setkeys{acs}{articletitle = true}
\SectionNumbersOn
\usepackage{amsmath}
\usepackage[nolist]{acronym}

\def\bra#1{\langle#1 |}
\def\ket#1{| #1 \rangle}

\usepackage{enumitem}

\usepackage{caption}
\usepackage{subcaption}
\usepackage{graphicx}
\usepackage{wrapfig}

\title{Reduced Scaling Real-Time Coupled Cluster Theory}
\author{Benjamin G. Peyton}
\affiliation{Department of Chemistry, Virginia Tech, Blacksburg, VA 24061, USA}
\author{Zhe Wang}
\affiliation{Department of Chemistry, Virginia Tech, Blacksburg, VA 24061, USA}
\author{T. Daniel Crawford}
\affiliation{Department of Chemistry, Virginia Tech, Blacksburg, VA 24061, USA}
\email{crawdad@vt.edu}
\begin{document}
\begin{acronym}
  \acro{LGPL}{GNU Lesser General Public Licence}
  \acro{WF}{wave function}
  \acro{WFT}{wave function theory}
  \acro{LHS}{left-hand side}
  \acro{CC}{coupled-cluster}
  \acro{CCS}{coupled-cluster with single substitutions}
  \acro{CCSD}{coupled-cluster with single and double substitutions}
  \acro{CCSD(T)}{CCSD with approximate triples correction}
  \acro{CCSDT}{coupled-cluster with single, double and triple substitutions}
  \acro{CC2}{approximate coupled-cluster singles and doubles}
  \acro{CC3}{approximate coupled-cluster singles, doubles and triples}
  \acro{EOM-CCSD}{equation-of-motion coupled-cluster singles and doubles}
  \acro{PT}{perturbation theory}
  \acro{MBPT}{many-body perturbation theory}
  \acro{MCSCF}{multiconfigurational self-consistent field}
  \acro{TDHF}{time-dependent Hartree-Fock}
  \acro{RPA}{random phase approximation}
  \acro{TDDFT}{time-dependent density-functional theory}
  \acro{DFT}{density-functional theory}
  \acro{SOPPA}{second-order polarization propagator}
  \acro{ADC}{algebraic diagrammatic construction}
  \acro{SCF}{self-consistent field}
  \acro{MO}{molecular orbital}
  \acro{NO}{natural orbital}
  \acro{CMO}{canonical molecular orbital}
  \acro{LMO}{localized molecular orbital}
  \acro{AO}{atomic orbital}
  \acro{MP}{M{\o}ller--Plesset}
  \acro{HF}{Hartree--Fock}
  \acro{QM}{quantum mechanics}
  \acro{QC}{quantum chemistry}
  \acro{RHS}{right-hand side}
  \acro{SCF}{self-consistent field}
  \acro{BCH}{Baker--Campbell--Hausdorff}
  \acro{MC}{Monte Carlo}
  \acro{FCI}{full configuration interaction}
  \acro{DMC}{diffusion Monte Carlo}
  \acro{QMC}{quantum Monte Carlo}
  \acro{VMC}{variational Monte Carlo}
  \acro{RI}{resolution of the identity}
  \acro{FCIQMC}{full configuration interaction quantum Monte Carlo}
  \acro{CCMC}{coupled cluster Monte Carlo}
  \acro{SCCT}{stochastic coupled cluster theory}
  \acro{RDM}{reduced density matrix}
  \acro{CI}{configuration interaction}
  \acro{FNA}{fixed-node approximation}
  \acro{LR}{linear response}
  \acro{ECD}{electronic circular dichroism}
\end{acronym}


\section*{Abstract} \label{abstract}
Real-time coupled cluster (CC) methods have several advantages over their
frequency-domain counterparts, namely, response and equation of motion CC
theories. Broadband spectra, strong fields, and pulse manipulation allow
for the simulation of complex spectroscopies which are unreachable using
frequency-domain approaches. Due to the high-order polynomial scaling,
the required numerical time-propagation of the CC residual expressions
is a computationally demanding process. This scaling may be reduced by
local correlation schemes, which aim to reduce the size of the (virtual)
orbital space by truncating it according to user-defined parameters. We
present the first application of local correlation to real-time CC. As in
previous studies of locally correlated frequency-domain CC, traditional
local correlation schemes are of limited utility for field-dependent
properties; however, a perturbation-aware scheme proves promising. A
detailed analysis of the amplitude dynamics suggests the main challenge
is a strong time-dependence of the wave function sparsity.


\section{Introduction}
\label{se:intro} 

Dynamic molecular responses induced by electromagnetic fields give rise to a number of experimental techniques for the
detailed investigation and characterization of light-matter interactions and structure,\cite{Barron2004} including
absorption, circular dichroism (CD), and Raman scattering spectroscopies.  Such techniques are essential for modern
synthetic chemistry in both research and industrial settings.  Theoretical/computational simluations of such spectra
have advanced to the point that they are now considered a ``full partner with experiment,''\cite{Goddard1985} providing
high-quality benchmark data for affirming or predicting many molecular responses, thereby increasing the accuracy and
reliability of spectral assignments and molecular structure determination. Computation of spectroscopic properties with
current \textit{ab initio} methods generally involves frequency-domain perturbation theory, commonly
referred to as response theory.\cite{Crawford2006,Norman2011,Helgaker2012}  

The development of modern response methods covers the broad range of quantum chemical techniques, including
\ac{TDHF}\cite{McLachlan64,Jorgensen75,Oddershede78,Jorgensen81} (also known as the \ac{RPA}\cite{Oddershede78}),
    \ac{SOPPA}\cite{Linderberg04}, the \ac{ADC}\cite{Schirmer82,Wormit14,Dreuw15}, \ac{MCSCF} theory\cite{Olsen85},
    \ac{TDDFT}\cite{Casida95}, M\o ller-Plesset perturbation theory\cite{Rice91,Aiga96}, and others.  (See the review by
            Norman\cite{Norman11} and the text by Norman, Ruud, and Saue\cite{Norman18} for an excellent overview of the
            many formulations and applications of response theory.) Coupled cluster (CC) response
    theory,\cite{Koch1990,Pedersen1997} in particular, has emerged as one of the most robust approaches to
    frequency-domain property calculations.\cite{Crawford2018,Crawford2019}

There are several drawbacks to the response formalism.\cite{Langhoff,Goings2018,Li2020} First and foremost, the
perturbations must be ``small'' relative to the intramolecular forces present in the system. This immediately precludes
the possibility of simulating high-energy experiments such as X-ray spectroscopy, which have numerous applications in
materials science, for example. Second, response theory
typically assumes weak, adiabatically-switched-on, monochromatic fields.
Experimental measurements, on the other
hand, can make use of complex, multi-phase procedures involving tuned laser pulses, pump-probe analysis,
    \textit{etc}.\cite{Maiuri2020} Finally, temporally controlled multi-photon events such as high harmonic generation
    (HHG)\cite{Lewenstein1994,Gorlach} lie outside the realm of the response formalism. Together, these drawbacks mean a
    wide variety of experiments cannot be straightforwardly predicted or supplemented with response theory calculations.
    To overcome this obstacle, it is useful to pursue non-perturbative, time-domain electronic structure theory,\cite{Goings2018,Crawford2019,Li2020}
    where there are fewer limitations on the form of the perturbing electromagnatic field.

Real-time time-dependent density functional theory
(RT-TDDFT) was introduced in the 1990s (then called the \textit{time-dependent local-density approximation}) 
\cite{Yabana1996,Yabana1997,Yabana1999,Bertsch2000} and provides a cost-effective approach to time-dependent
simulations.\cite{Lopata2011,Castro2015,Tussupbayev2015,Goings2016a,Bruner2016,Goings2018,Sun2019a,Li2020}
Efforts to introduce computational savings into RT-TDDFT have largely focused on reducing
simulation times, utilizing techniques such as Pad\'e approximants to
accelerate the convergence of the Fourier transform\cite{Bruner2016}
and fitting schemes to avoid the Fourier transform all together,
thus ameliorating the problem of short trajectories resulting in low-resolution
spectra.\cite{Ding2013} Repisky \textit{et al.} introduced the concept of 
dipole-pair contributions,\cite{Repisky2015,Kadek2015}
which are typically less complicated than the total electric dipole,
and so these may be individually approximated efficiently using the 
techniques mentioned above.\cite{Bruner2016} However, the challenges of frequency-domain DFT carry
over directly to the time domain, such as the underestimation of excited
state energies\cite{Peach2008} and difficulties arising from the adiabatic
approximation.\cite{Fuks2013,Fuks2015,Bruner2016} We refer the reader to
a recent, comprehensive review article\cite{Li2020} and citations therein
for a more complete discussion of these challenges. Regardless, the success
of RT-TDDFT under most conditions combined with its drastically reduced
computational cost make it the only viable method for large molecules
at present.

The alternative of real-time CC (RTCC) methods has been discussed nearly as far back as the origins of CC itself in the
realm of nuclear physics.\cite{Hoodbhoy1978,Hoodbhoy1979,Gunnarsson78} More recently, a renewed interest in real-time
coupled cluster has developed for the reasons discussed above, and in the past ten years, several implementations have
been reported, \cite{Huber2011,Kvaal2012,Nascimento2019,Pedersen2019,Park2019} with new insights into the aspects of
numerical integration\cite{Pedersen2019,Kristiansen2020} and interpretation\cite{Pedersen2019,Pedersen2021} as well as
applications for a number of spectral properties.\cite{Nascimento2016,Nascimento2017,Nascimento2019,Park2019,Park2021b}
Orbital adaptive\cite{Kvaal2012} and orbital optimized\cite{Sato2018} variants have also explored the limitations of
unrelaxed canonical Hartree Fock orbitals, as well as the effects of alternative reference orbitals on the propagation
of unphysical imaginary components to energetics and electric dipole moments.  Notably absent, however, are studies on
the ability to reduce the cost of real-time coupled cluster methods. 

Borrowing from the vast literature of reduced-scaling ground-state or frequency-domain CC, there are numerous potential
candidates for reducing the cost of RTCC, beyond the successful approaches implemented for RT-TDDFT described above.
First, the standard non-perturbative truncated approaches used for properties such as CC2\cite{Christiansen1995} and
CC3\cite{Koch1997} are immediately possible, as are property-optimized basis sets.
\cite{Wolinski1990,Sadlej1977,Roos1985,Sadlej1991a,Benkova2005,Baranowska2010,Baranowska2013,Aharon2020a,Howard2018}
Further, details of implementation such as the choice of intermediate tensors, the effects of single- or mixed-precision, or
the use of graphical processing units have only just begun to be explored.\cite{Wang2022} An alternative formulation
developed separately by the DePrince and Bartlett groups, dubbed the time-dependent equation-of-motion CC (TD-EOM-CC)
    method,\cite{Nascimento2016,Nascimento2017,Nascimento2019,Park2019,Park2021b} reduces the cost by targeting the
    difficulty of numerical integration of multiple ``stiff'' coupled differential equations. By selecting a given
    moment function to propagate in time, the coupled right- and left-hand wave function amplitudes of CC theory do not
    have to be propagated separately, reducing both the number and difficulty of numerical integrations required. 

The principal drawback for RTCC theory, however, is its high-degree polynomial scaling, which further increases the cost
of even short propagations in time.  For time-independent ground-state CC energy calculations, numerous techniques --- 
referred to as ``local correlation'' or ``reduced-scaling'' methods --- have
been developed over the last few decades to ameliorate this shortcoming.\cite{Pulay83, Stoll92, Saebo93, Hampel96,
    Schutz99, Schutz00:Ta, Schutz00:Tb, Schutz01, Schutz02, Mata07, Taube08, Piecuch09, Neese09, Neese09:CEPA,
    Piecuch10, Piecuch10b, Liakos11, Tew11, Yang11, Yang12, Liakos12, Hattig12, Riplinger13, Krause12, Masur13,
    Pavosevic2014-xg, Werner15, Pavosevic2016-kw, Ma17, Pavosevic17} These methods seek to build a compact
    virtual-orbital (correlation) space based on lower-cost criteria, such as (pair) energies from low-order
    perturbation theory or atomic-orbital charge analysis thereby reducing the exponential cost of the method, possibly
    as far as linear scaling. While still only routine for ground-state calculations, these methods and variants thereof
    have shown promise in the calculation of selected response properties. \cite{McAlexander2016,Kumar2017,Howard2018,DCunha2021,DCunha2022}

In this work, we report the first application of local correlation to RTCC.  The effects of occupied and virtual space
localization are considered for the simulations of small hydrogen clusters in the presence of electric field
perturbations. Absorption cross sections as well as electronic CD (ECD) spectra are computed using successively smaller
fractions of the canonical orbital space using the popular projected atomic orbital (PAO)
    \cite{Pulay1983,Saebo1985,Saebo1986,Saebo1993} and pair natural orbital (PNO) \cite{Neese2009,Neese2009a} schemes.
    Additionally, we compare these to the recent perturbed pair natural orbital (PNO++) approach,
    \cite{Crawford2019,DCunha2021,DCunha2022} which has been formulated specifically for computing field-induced
    properties.  The results are analyzed with respect to full-space simulations.  Finally, wave function amplitude
    dynamics are investigated in order to determine the extent to which these schemes suppress or cause large amplitude
    deviations, which cause instabilities in numerical integration and spurious oscillations in the dipole trajectory.


\section{Theoretical Background} \label{se:theory}
\subsection{Real-Time Coupled Cluster Theory} \label{ss:rtcc}
Conventional RTCC implementations begin by solving the ground-state right- and left-hand
CC wave function amplitude (residual) equations in the absence of the external field, \textit{viz.}, 
\begin{subequations}
    \begin{equation} \label{eq:t_res}
        \bra{\mu}\bar{H}\ket{\Phi} = 0
    \end{equation}
    \begin{equation} \label{eq:l_res}
        \bra{\Phi}(1 + \hat\Lambda)[\bar{H},\tau_\mu]\ket{\Phi} = 0
    \end{equation}
\end{subequations}
where $\ket{\Phi}$ is the Hartree-Fock ground state determinant and $\ket{\mu}$ 
are substituted determinants (singles, doubles, \textit{etc}.) obtained by the 
action of the second-quantized excitation and de-excitation operators $\tau_\mu$,
and $\bar{H}$ is the similarity transformed electronic Hamiltonian
\begin{equation} \label{eq:hbar}
    \bar{H} = e^{-\hat{T}}\hat{H}e^{\hat{T}}
\end{equation}
with right-hand cluster operators 
\begin{subequations}
    \begin{equation}
        \hat{T} = \sum_\mu^N \hat{T}_\mu
    \end{equation}
    \begin{equation} \label{eq:t_mu}
        \hat{T}_\mu = \sum_{\mu}\tau_\mu t_{\mu}
    \end{equation}
\end{subequations}
and left-hand cluster operators
\begin{subequations}
    \begin{equation}
        \hat{\Lambda} = \sum_\mu^N \hat{\Lambda}_\mu
    \end{equation}
    \begin{equation} \label{eq:l_mu}
        \hat{\Lambda}_\mu = \sum_{\mu}\tau_\mu^\dagger\lambda_\mu.
    \end{equation}
\end{subequations}
The time evolution of the amplitudes is governed by the nonlinear differential equations
obtained through the time-dependent Schr\"odinger equation (in atomic units)
\begin{subequations}
    \begin{equation} \label{eq:diff_t}
        i\frac{d t_\mu}{d {t}} = \bra{\mu}\bar{H}(t)\ket{\Phi}
    \end{equation}
    \begin{equation} \label{eq:diff_l}
        -i\frac{d \lambda_\mu(t)}{d{t}} = \bra{\Phi}(1 + \hat\Lambda(t))[\bar{H}(t),\tau_\mu]\ket{\Phi}.
    \end{equation}
\end{subequations}
The right-hand sides of Eqs.~(\ref{eq:diff_t}) and (\ref{eq:diff_l}) are
the same as the residual
expressions above, where we have explicitly noted the time dependence of
the Hamiltonian and the cluster operators. By 
including a field perturbation as a time-dependent addition to the Fock operator, the right-hand
sides of these equations may be computed by updating the Hamiltonian from time $t$ to time $t^\prime = t + h$ and 
recomputing the residual expressions. 
This is achieved using a numerical integrator, which produces solutions to equations 
of the form 
\begin{equation}
    \frac{dy(t)}{dt} = f(y,t).
    \label{eq:ode}
\end{equation}
Here, $y$ is a concatenation of the $t_\mu$ and $\lambda_\mu$ vectors, and
the function $f$ represents a corresponding concatenation of the residual
expressions. 
Multiple integration schemes are possible, and for simplicity we adopt the popular
fourth-order Runge-Kutta integrator,\cite{rk} defined by 
\begin{equation}
\begin{aligned}
    k_1 &= f\left(y,t\right) \\
    k_2 &= f\left(y+\frac{1}{2}hk_1,t+\frac{1}{2}h\right) \\
    k_3 &= f\left(y+\frac{1}{2}hk_2,t+\frac{1}{2}h\right) \\
    k_4 &= f\left(y+hk_3,t+h\right)
\end{aligned}
\end{equation}
with time step $h$, and the propagated tensor is computed as
\begin{equation}
    y(t+h) = y(t) + \frac{1}{6}h(k_1 + 2k_2 + 2k_3 + k4).
\end{equation}

\subsection{Properties} \label{ss:prop}
Within the dipole approximation, 
the complex time-domain response tensors 
$\boldsymbol{\alpha}$ (the polarizability tensor) and 
$\textbf{G}^\prime$ (the optical activity/Rosenfeld tensor) can be defined by low-order expansions 
of the time-dependent electric and magnetic dipole moment expectation values 
in an electric field $\textbf{E}$ with frequency $\omega$, \textit{viz.}
\begin{subequations} \label{eq:exps}
    \begin{equation} \label{eq:mu_exp}
        \langle\mu\rangle_i = \mu_0 + \alpha_{ij}E_j
    \end{equation}
    \begin{equation} \label{eq:m_exp}
        \langle m\rangle_i = -\frac{1}{\omega}\frac{\partial E_j}{\partial t}G^\prime_{ij}
    \end{equation}
\end{subequations}
where $i$ and $j$ are Cartesian coordinates, and the notation suppresses
the time dependence for clarity.  
The dipole strength function $S$
is related to the 
imaginary component of $\boldsymbol{\alpha}$ by
\begin{equation} \label{eq:abs}
    S(t) \propto \textrm{Tr}\left[ \textrm{Im}\left( \boldsymbol{\alpha}(t) \right) \right], 
\end{equation}
and Fourier transformation of $S$ to the frequency domain
yields the broadband absorption spectrum. 
The differential molar extinction coefficient is proportional to the imaginary part
of the Rosenfeld $\textbf{G}^\prime$ tensor\cite{Rosenfeld1929} by
\begin{equation} \label{eq:ecd}
    \eta(t) \propto -\textrm{Tr}[\textrm{Im}(\textbf{G}^\prime(t))],
\end{equation}
and the Fourier transform of Eq.~(\ref{eq:ecd}) yields the \ac{ECD} spectrum.

We note here two important points. First, we could just as easily
define both $\boldsymbol{\alpha}$ and $\textbf{G}^\prime$ with respect
to the electric dipole expectation value; however, by expanding both
moments in an electric field, we may recover both properties by computing
expectation values of both the electric and magnetic dipole operators
along the same electric field-perturbed trajectory. In principle, we may
compute \textit{any} electric-field-perturbed expectation value from a
single propagation -- this is in contrast to the RT-EOM-CC method, which
propagates a single moment function. Additional expectation values
would require additional moment function propagations.


Second, we note that the low-order expansions in Eq.~(\ref{eq:exps}) are
an approximation. The total dipole moments will contain many higher-order
terms; however, at the field strengths used in this work, these effects
are expected to be negligible. These terms can be separated and have been
examined in the context of real-time simulations of X-ray absorption 
spectroscopy (XAS).\cite{Park2021b}
While very important to the advantages of the RTCC method, these effects
are beyond the scope of the current work.

\subsection{Local Correlation} \label{ss:local}
\subsubsection{Projected Atomic Orbitals} \label{sss:pao}

In the PAO method, the virtual space is localized based on spatial criteria derived from the set of atom-centered atomic
orbital (AO) basis functions.  The occupied orbitals are first localized using conventional criteria such as that defined
by Pipek and Mezey\cite{Pipek1989} or by Boys.\cite{Boys60}  Then, the virtual space is localized by projecting away
contributions from the occupied space,
\begin{equation}
    \tilde{\mathbf{C}}^\prime = \mathbf{1} - \mathbf{DS},
\end{equation}
where $\mathbf{D}$ is the Hartree-Fock density matrix and $\mathbf{S}$ is the overlap (both in the AO basis), and
$\tilde{\mathbf{C}}^\prime$ is the resulting PAO coefficient matrix, which has the same dimension as the original AO basis.
Thus, the PAOs are orthogonal to the occupied space, but remain non-orthogonal among themselves.  

The reduction of the size of the correlation (virtual) space is achieved
by choosing a subset of the PAOs --- the virtual ``domain'' --- for each
occupied orbital.  The selection process is based on an atom-by-atom
approach in which all the AO basis functions associated with a given
atom are used to compute a ``completeness function'' for each occupied
orbital,\cite{Boughton93} 
\begin{equation}
    b_i(\tilde{\mathbf{C}}^\prime) = 1 - \sum_{\mu \in i}\sum_{\lambda}C^{\prime}_{\mu i}S_{\mu\lambda}C_{\lambda i},
\end{equation} 
where $i$ labels the given occupied MO and $\mathbf{C}$ is the
original MO coefficient matrix. The left-hand sum is limited to those AOs
currently contained within the domain of MO $i$, such that 
$C^{\prime}_{\mu i}$ is the PAO MO coefficient matrix defined on the domain associated
with the current iteration.  If the chosen subset of atoms and their AOs
do not yield a value of $b_i$ below an assigned cutoff, $\delta_{PAO}$,
the additional atoms are added in a ranked order (commonly based on
atomic Mulliken populations/charges) until the value of $b_i$
converges.  Once the domains are assigned, occupied
\textit{pairs} $ij$ are then assigned \textit{pair} domains, based on
the union of the domains of the two occupied orbitals, and the number of
such pairs included in a given computation is often subject to distance-
or energy-based critera.  A key aspect of the PAO approach is that each
occupied orbital (or orbital pair) uses a virtual/correlation domain that
is taken from a common set of PAOs.

The PAO space contains as many linear dependencies as occupied orbitals, and these are removed by diagonalizing the projected 
overlap matrix
\begin{equation}
    \tilde{\mathbf{S}}_{ij} = \tilde{\mathbf{C}}^{\prime \dagger}_{ij}\mathbf{S}\tilde{\mathbf{C}}^{\prime}_{ij},
\end{equation}
where $\tilde{\mathbf{C}}^\prime_{ij}$ contains only the columns for atomic orbitals belonging to
the domain of pair ${ij}$. PAOs that correspond to eigenvalues of $\tilde{\mathbf{S}}_{ij}$ 
below a cutoff parameter are removed, and the remaining orbitals are normalized to yield the non-redundant PAO basis for
a given pair, $\tilde{\mathbf{C}}_{ij}$.

Two matrices are required to transform MO-basis quantities 
into the PAO basis. The first, which rotates from the MO to the redundant PAO
basis for a given pair, is computed as
\begin{equation} \label{eq:Q_pao}
    \mathbf{Q}^{PAO}_{ij} = \mathbf{C}^{\dagger}_{ij}\mathbf{S}\tilde{\mathbf{C}}^\prime_{ij}.
\end{equation}
To facilitate the use of the usual orbital energy denominator terms during the update of the 
amplitude equations at every iteration, a semi-canonical virtual basis for pair $ij$,
$L^{PAO}_{ij} = \chi_{ij}\tilde{C}_{ij}$,
is found by diagonalizing the Fock matrix in the space of non-redundant PAOs $\tilde{F}$:
\begin{equation} \label{eq:L_pao}
    \tilde{F}\chi_{ij} = \epsilon_{ij}\chi_{ij}
\end{equation}
where $\epsilon_{ij}$ are the semi-canonical orbital energies for the virtual space of 
occupied pair $ij$. 

\subsubsection{Pair Natural Orbitals}

In the PNO approach, a compact virtual space is reduced based on density-based criteria rather than spatial extent.
Just as in the PAO method, the occupied orbitals are first localized, but then the virtual orbitals are obtained though
diagonalization of an approximate density defined in the canonical virtual space for each occupied pair, $ij$.  The most
common choice for this density is that obtained from second-order M{\o}ller-Plesset perturbation theory (MP2), viz.,
\begin{equation} \label{eq:pair_D}
    \mathbf{D}_{ij} = \frac{2}{1+\delta_{ij}}(\mathbf{T}_{ij}\tilde{\mathbf{T}}_{ij}^\dagger +
            \mathbf{T}_{ij}^\dagger\tilde{\mathbf{T}}_{ij}),
\end{equation}
where $\mathbf{T}_{ij}$ is the matrix of first-order double-excitation amplitudes, and $\tilde{\mathbf{T}}_{ij} = 2\mathbf{T}_{ij} - \mathbf{T}_{ij}^\dagger$.
Diagonalizing $\mathbf{D}_{ij}$ yields the transformation matrix from the virtual MO to the PNO basis,
\begin{equation} \label{eq:Q_pno}
    \mathbf{D}_{ij} \mathbf{Q}^{PNO}_{ij} = \mathbf{Q}_{ij} \mathbf{n}_{ij}
\end{equation}
where $\mathbf{n}_{ij}$ is the diagonal matrix of occupation numbers. Truncation of the space is accomplished by removing 
PNOs corresponding to occupation numbers below a cutoff, $\delta_{PNO}$. Similarly to the PAO approach, the transformation
matrix from the PNO to a semi-canonical PNO basis for each pair is obtained by diagonalizing the virtual-virtual block of the Fock matrix
in the space of PNOs:
\begin{equation} \label{eq:L_pno}
    \tilde{\mathbf{F}}\mathbf{L}^{PNO}_{ij} = \mathbf{L}^{PNO}_{ij} \boldsymbol{\epsilon}_{ij}.
\end{equation}

A variant of the PNO space developed by Crawford $\textit{et al.}$, known as perturbed PNOs or PNO++,
has been applied to linear and quadratic response properties in the frequency domain.\cite{Crawford2019,DCunha2021,DCunha2022,Kodrycka2022} To improve
the ability of the pair density to estimate the relative importance of PNOs to field-dependent
quantities, the approximate amplitudes $\mathbf{T}_{ij}$ in Eq.~(\ref{eq:pair_D}) are replaced by
approximate \textit{perturbed} amplitudes $\mathbf{X}^{ij}_B$
\begin{equation} \label{eq:pert_D}
    \mathbf{D}_{ij}(B) = \frac{2}{1+\delta_{ij}}(\mathbf{X}_B^{ij}\tilde{\mathbf{X}}_B^{ij\dagger}
    + \mathbf{X}_B^{ij\dagger}\tilde{\mathbf{X}}_B^{ij})
\end{equation}
with approximate MP2-level perturbed amplitudes created using the Hamiltonian and
perturbation $B$ (chosen to be the electric dipole moment in the length representation) operators, 
similarity transformed according to Eq.~(\ref{eq:hbar}),
\begin{equation} \label{eq:pert_amp}
    X^{ij}_{ab} = \frac{\bar{B}}{\bar{H}_{aa} + \bar{H}_{bb} - \bar{H}_{ii} - \bar{H}_{jj}}.
\end{equation}
We note here that Ref.~\citenum{DCunha2021} includes a factor of the field frequency in the 
denominator in Eq.~(\ref{eq:pert_amp}). As this is generally chosen to be the frequency at which 
the property is measured, and given the broadband nature of the target absorbance and 
ECD spectra, we have chosen to omit this term. 
Once created, the perturbed pair density may be used to obtain the PNO++ space in an identical
manner to the PNO procedure, which is then truncated using a cutoff $\delta_{PNO++}$.

\section{Computational Details} \label{se:comp}

We propagated the CC wave function for a helical H$_2$ tetramer [atomic coordinates provided in the supplementary
information (SI)] for 800 a.u.\ using a time step of 0.02 a.u. in the presence of a time-dependent electric field. 
In order to engage all excited states, we approximated a Dirac delta pulse as a narrow Gaussian field 
of the form
\begin{equation}
    E(t) = \textrm{F}e^{-\frac{(t-\nu)^{2}}{2\sigma^2}},
\end{equation}
with field strength $\textrm{F}$, pulse center $\nu$, and standard deviation $\sigma$.
The field was oriented in the y-direction, which was along the helical axis of the system.
All calculations in this work used a field defined by $\textrm{F} = 1\times 10^{-3}$,
$\nu = 0.05$, and $\sigma = 0.01$, all in atomic units. Time-dependent electric and magnetic 
dipole moments were damped using a function of the form $e^{-t\tau}$, 
with $\tau = 150$.
The cc-pVDZ basis
set augmented with diffuse functions\cite{Dunning1989,Woon1994} was used throughout.  

The reference, PAO, and PNO simulations were performed in the MO space following localization of the occupied
orbitals using the Pipek-Mezey procedure.\cite{Pipek1989}  The PAO and PNO simulations
were carried out using domain cutoffs
corresponding to successively smaller fractions of the MO virtual space. 
The relative sizes of the remaining virtual spaces are measured by the $T_2$ ratio,
defined as the ratio of the number of doubles amplitudes in the truncated and
untruncated spaces. Lower $T_2$ ratios indicate
more aggressive truncation of the virtual space, resulting in potential computational savings.
$T_2$ ratios from $0.05$ to $0.95$ were explored for each localization scheme.

The effect of truncation of the virtual space was computed using a modified version of our canonical-MO CC
code as first described by Hampel and Werner\cite{Hampel1996} and used in our previous work in reduced-scaling CC
methods.\cite{Crawford02,McAlexander2016,DCunha2021,DCunha2022}
In each CC iteration or time-propagation step, the amplitude residuals given by Eqs.~(\ref{eq:t_res}) and (\ref{eq:l_res})
are transformed into the PAO or PNO representation using $\mathbf{Q}$ and into the semi-canonical
basis using $\mathbf{L}$ as
\begin{subequations} \label{eq:rotate}
\begin{equation} \label{eq:rotate_r1}
    \tilde{\textbf{r}}_i = \textbf{L}_{ii}^T\textbf{Q}_{ii}^T\textbf{r}_i
\end{equation}
and
\begin{equation} \label{eq:rotate_r2}
    \tilde{\textbf{r}}_{ij} = \textbf{L}_{ij}^T\textbf{Q}_{in}^T\textbf{r}_{ij}\textbf{Q}_{in}\textbf{L}_{ij}.
\end{equation}
\end{subequations}
The transformation matrices are computed from either Eqs.~(\ref{eq:Q_pno}) 
and (\ref{eq:L_pno}), respectively, for PNOs, or Eqs.~(\ref{eq:Q_pao}) and (\ref{eq:L_pao}),
respectively, for PAOs.  
For the ground-state (time-independent) residuals, the energy denominator is applied in this
basis and then the resulting amplitudes are back-transformed into the MO basis, but for the time-dependent residuals, no
such denominators are required, in accord with Eq.~(\ref{eq:ode}).  These transformations yield MO-basis amplitudes for
which the non-local components have been eliminated.

For absorption spectra, 
the imaginary component of the Fourier transform of the \textit{induced} electric dipole 
$\tilde{\mu} = (\langle\mu\rangle - \mu_0)$ following an electric-field kick 
may be directly divided by the Fourier transform of the field strength to yield the spectrum. 
Electronic CD spectra, however, require the Fourier transform of the time-derivative of the field.
For a Dirac delta pulse $E_\delta(t) = \kappa\delta(t)$, 
the Fourier transform of the derivative yields
\begin{equation}
    \textrm{FFT}[E_\delta] = i\omega\kappa.
\end{equation}
Therefore, for such a field, the CD is proportional to the negative of the 
\textit{real} part of the Fourier 
transform of the induced magnetic dipole. 
In practice, the assumption of a Dirac delta pulse is sufficient, provided
a thin Gaussian or Lorentzian pulse is used. 

Discrete Fourier transformation was done using a wrapper to the \texttt{fft} submodule
of the SciPy python library.\cite{scipy} 
All methods were implemented in the Python-based coupled
cluster package, PyCC\cite{pycc}, a NumPy-based\cite{numpy} open-source code developed 
in the Crawford group for the testing and implementation of novel coupled cluster methods. 
The code utilizes the Psi4 electronic
structure package\cite{Smith2020} for integral generation and computing reference 
wave functions. 
The RTCC code makes use of the \texttt{opt\_einsum} package\cite{opteinsum} for tensor contractions,
and time propagation is performed using an in-house suite of integrators. The integrator
used throughout this work was the fourth-order Runge-Kutta method.\cite{rk}


\section{Results and Discussion} \label{se:results} 

Here we present results from the first applications of local correlation 
methods to RTCC. In Section
\ref{ss:spectra}, we examine the convergence of absorption and ECD spectra for
a helical hydrogen-molecule cluster, as compared to canonical (untruncated)
RTCC simulations, followed by an analysis of the corresponding cluster
amplitude dynamics in Section \ref{ss:amps}.  In addition, we explore the
effect of localization and truncation on the orbital extent in an attempt to
explain apparent deficiencies of the PNO space in light of variations in the
time-dependent deviations in the amplitudes from the ground-state.

\subsection{Absorption and ECD Spectra} \label{ss:spectra}
\subsubsection{Absorption} \label{sss:abs}
\begin{figure} 
    \centering
    \includegraphics[scale=.8]{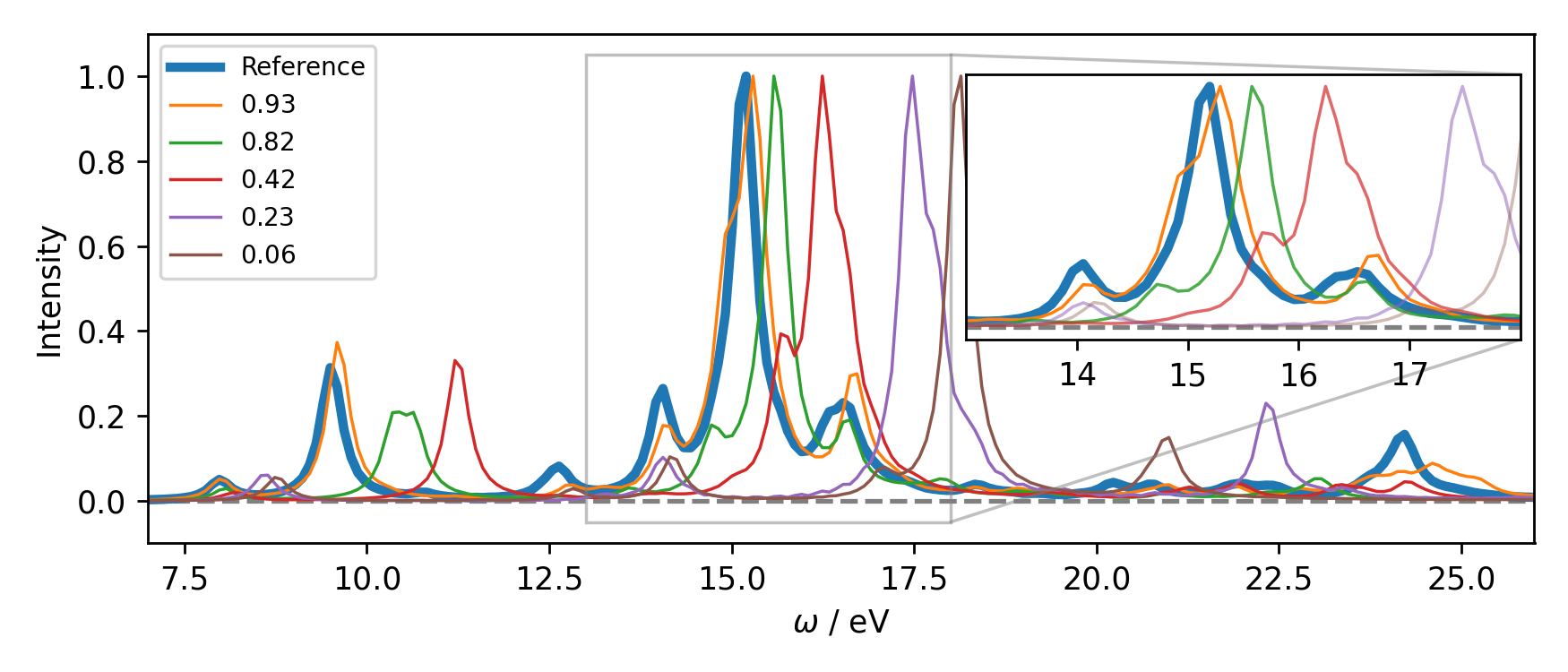}
    \caption{Reference and PAO absorption spectra of $(H_2)_4$ for five cutoffs: 
    [$1\times 10^{-4}$, $1\times 10^{-3}$, $1\times 10^{-2}$, $5\times 10^{-2}$, 
    $1\times 10^{-1}$] corresponding to $T_2$ ratios of 
    [$0.93$, $0.82$, $0.42$, $0.23$, $0.06$], respectively.}
    \label{fig:pao_abs}
\end{figure}

Absorption spectra are obtained from the Fourier transform of Eq.~(\ref{eq:abs}).
Fig.~\ref{fig:pao_abs} shows the normalized absorption spectrum obtained from
a reference propagation along with five PAO cutoffs resulting in $T_2$ ratios
ranging from $0.06$ to $0.93$. (The corresponding $\delta_{PAO}$ cutoffs are
included in the Figure caption.)

The largest truncated PAO virtual space, corresponding to a $T_2$ ratio of
$0.93$, accurately predicts the excitation energies for each major peak below
17 eV.  Convergence to the large reference peak near 15.5 eV occurs from the
right, indicating a lowering of excitation energies as the size of the virtual
space increases.  This is the expected trend as the size of the correlation
space for the time-dependent wave function gradually approaches the same
quality as the ground state (unperturbed) wave function and can also be
observed for the smaller peak near 10 eV.  However, the accuracy of these peaks
rapidly declines, e.g.\ at a ratio of $0.82$ where the base peak position is
off by approximately 0.5 eV.  Performance continues to degrade as the
excitation energy increases and the average size of the PAO space decreases.
For the final two cutoffs, at $T_2$ ratios of $0.21$ and $0.07$, the base
peaks are 3 eV or more away from the reference, and no peak is exhibited near
25 eV. These spaces also fail to predict the second largest peak, occuring
just below 10 eV.  Thus, while the 0.93 $T_2$ ratio ($\delta_{PAO} = 10^{-4}$)
reproduces the reference spectrum adequately for the lower-energy peaks, the
broadband laser pulse that excites all electronic states puts too great of a
burden on the global PAO truncation.

\begin{figure} 
    \centering
    \includegraphics[scale=.8]{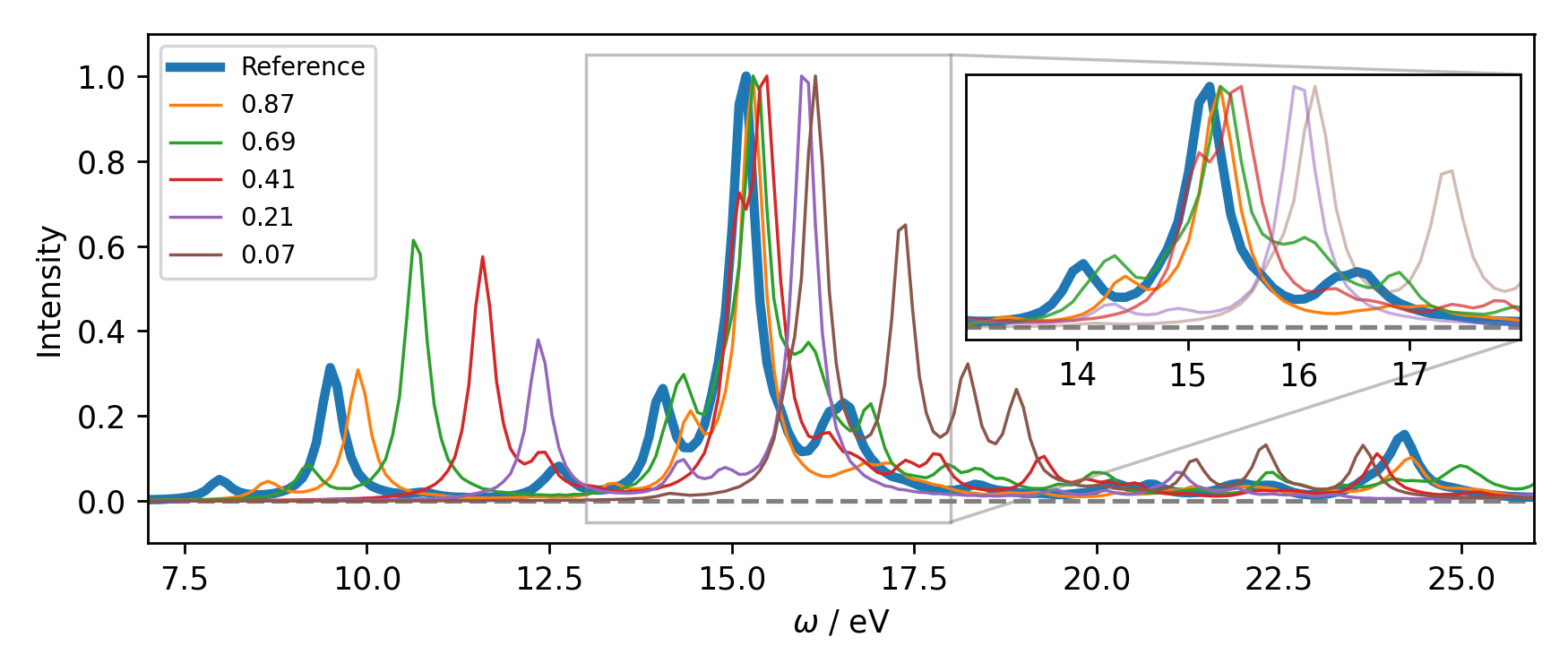}
    \caption{Reference and PNO absorption spectra of $(H_2)_4$ for five cutoffs: 
    [$1\times 10^{-10}$, $1\times 10^{-9}$, $1\times 10^{-8}$, $1\times 10^{-7}$, 
    $2\times 10^{-6}$] corresponding to $T_2$ ratios of 
    [$0.87$, $0.69$, $0.41$, $0.21$, $0.07$], respectively.}
    \label{fig:pno_abs}
\end{figure}

The performance of the PNO approach for the (H$_2$)$_4$ case is shown in Fig.~\ref{fig:pno_abs}. 

The truncated PNO virtual spaces overestimate the position of the base peak at
15.5 eV, with the smallest space ($T_2$ = 0.07) predicting a peak within 1.5 eV
of the reference and the two largest spaces ($T_2$ = 0.87 and 0.69)
predicting this peak to within 0.2 eV of the reference.  Even a PNO space with
a $T_2$ ratio of just $0.41$ predicts the base peak position with more
accuracy than a PAO space where the $T_2$ ratio has doubled.  However,
convergence of the shoulder peaks on either side of the base peak, indicated
by the inset of Fig.~\ref{fig:pno_abs}, is less predictable. Even the
largest spaces considered do not correctly reproduce the excitation energy for
these peaks, with no clear advantage to having a $T_2$ ratio of $0.87$ when
compared to a ratio of just $0.69$.  This trend continues into the
higher-energy range of the spectrum, with the performance of each cutoff being
nearly indistinguishable.

\begin{figure}
    \centering
    \includegraphics[scale=.8]{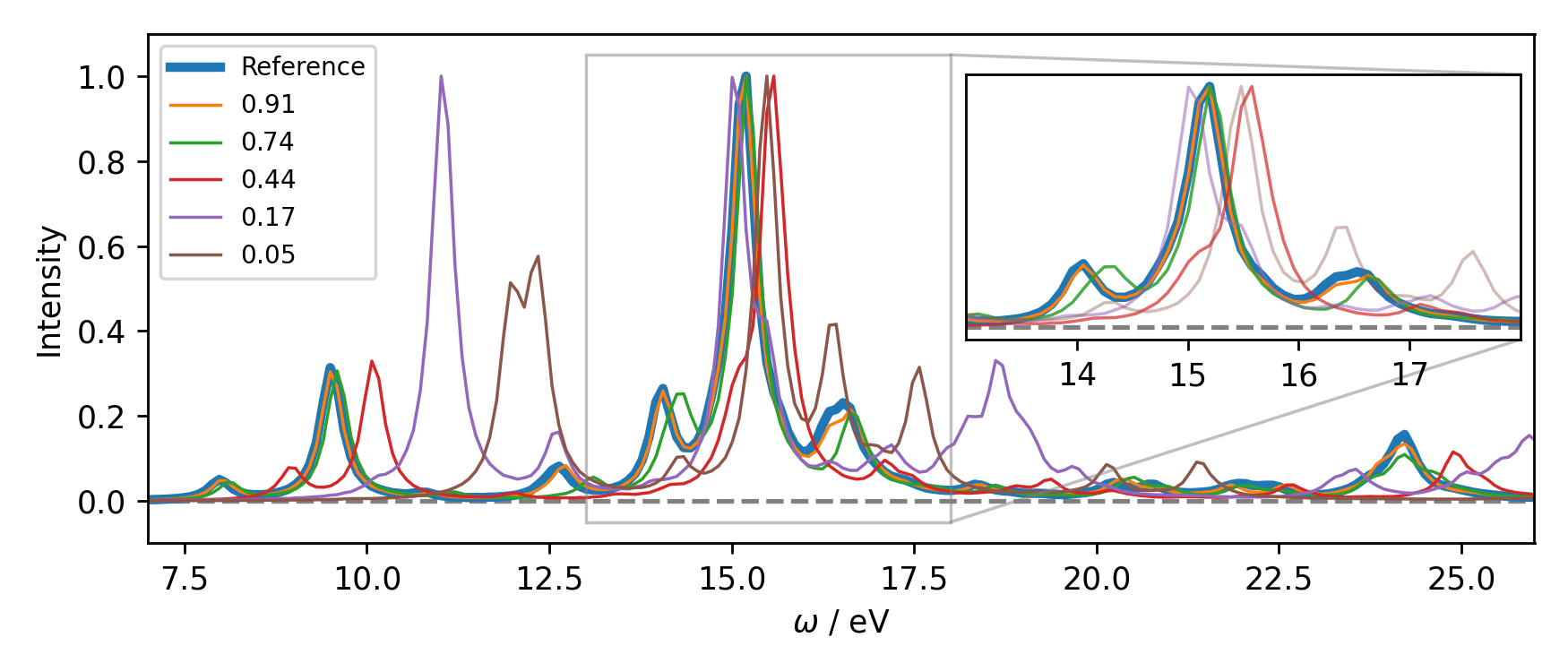}
    \caption{Reference and PNO++ absorption spectra of $(H_2)_4$ for five cutoffs: 
    [$1\times 10^{-9}$, $1\times 10^{-8}$, $1\times 10^{-7}$, 
    $1\times 10^{-6}$, $1\times 10^{-5}$] corresponding to $T_2$ ratios of 
    [$0.91$, $0.74$, $0.44$, $0.17$, $0.05$], respectively.}
    \label{fig:pnopp_abs}
\end{figure}

Overall, while the PNO basis exhibits improvements for several peaks over the PAO basis, 
neither scheme produces adequate results upon aggressive truncation of the virtual space.
This result is not entirely surprising; in studies of local correlation applied to 
response theory by Crawford \textit{et al}., 
\cite{McAlexander2016,Kumar2017,Crawford2019,DCunha2021,DCunha2022} 
traditional reduced-scaling approachs proved inaccurate for the closely related
electric-dipole polarizability.
The perturbation-aware PNO++ space is a promising alternative for such
calculations, with results
for the absorption spectrum shown in Fig.~\ref{fig:pnopp_abs}.

While aggressive truncation beyond a $T_2$ ratio of 0.5 is still not feasible,
larger PNO++ spaces far outperform their PAO or PNO counterparts. For example,
at a $T_2$ ratio of 0.74, PNO++ provides an improvement over even the largest
PAO and PNO spaces tested. Notably, the troublesome peak position near 12.5 eV
has improved, and is nearly overlapping the reference spectrum at a ratio of
0.91. This result corroborates those from frequency-domain studies, suggesting
the interchangeability of reduced-scaling methods from frequency- to
time-domain CC theory.

\subsubsection{ECD} \label{sss:ecd}

In terms of response theory, absorption is related to the \textit{imaginary}
part of the electric dipole--electric dipole linear response tensor
($\boldsymbol{\alpha}$ in Eq.~(\ref{eq:mu_exp})), while the scalar
polarizability and the refractive index are related to the \textit{real} part.
Indeed, all linear absorptive properties such as absorption and CD are related
to the imaginary component of a linear response tensor, while dispersive
properties such as refractive index and circular birefringence (also known as
optical rotation) are related to the real component.
\cite{Barron2004,Norman2011} 

\begin{figure} 
    \centering
    \includegraphics[scale=.8]{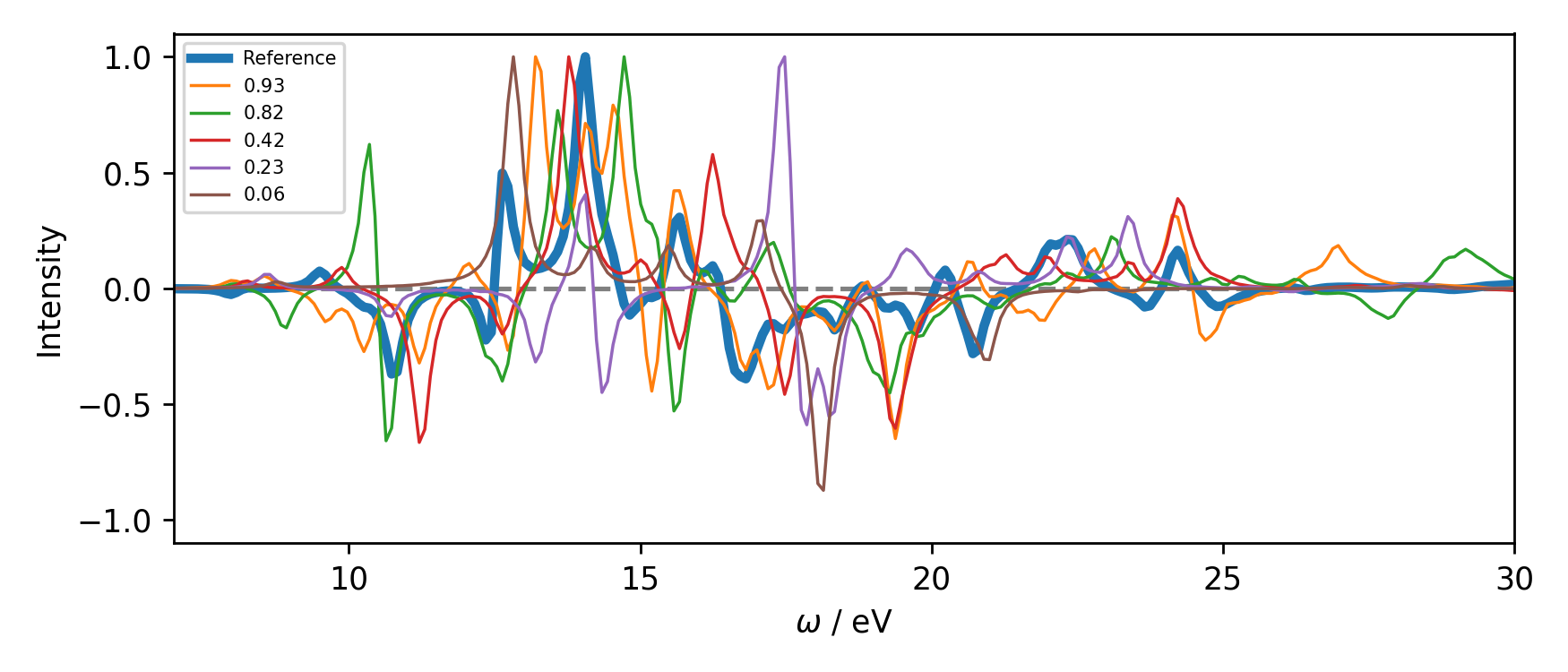}
    \caption{Reference and PAO ECD spectra of $(H_2)_4$ for five cutoffs: 
    [$1\times 10^{-4}$, $1\times 10^{-3}$, $1\times 10^{-2}$, $5\times 10^{-2}$, 
    $1\times 10^{-1}$] corresponding to $T_2$ ratios of 
    [$0.93$, $0.82$, $0.42$, $0.23$, $0.06$], respectively.}
    \label{fig:pao_ecd}
\end{figure}

The ECD spectrum is obtained from the Fourier transform of Eq.~(\ref{eq:ecd}).
Being a bisignate, mixed-response property, ECD is a considerable computational
challenge, similar to its dispersive counterpart circular birefringence. 
Fig.~\ref{fig:pao_ecd} shows the results for an ECD spectrum in the same PAO 
orbital spaces used for the absorption spectrum in the previous section.

The dynamic response of the magnetic dipole to the electric field in this
frequency range is considerably more complicated than that of the electric
dipole. With even the largest spaces tested, virtually all distinguishing
characteristics of the reference spectrum are unidentifiable, and the
performance of the PAO basis near the base peak 
varies wildly with truncation.
Curiously, the two largest spaces predict significant peaks above 25 eV that
are not present in the reference or the smaller PAO spaces. This indicates a
strong sensitivity of the response of the wave function to the completeness
threshold used for determining the occupied domains.

\begin{figure} 
    \centering
    \includegraphics[scale=.8]{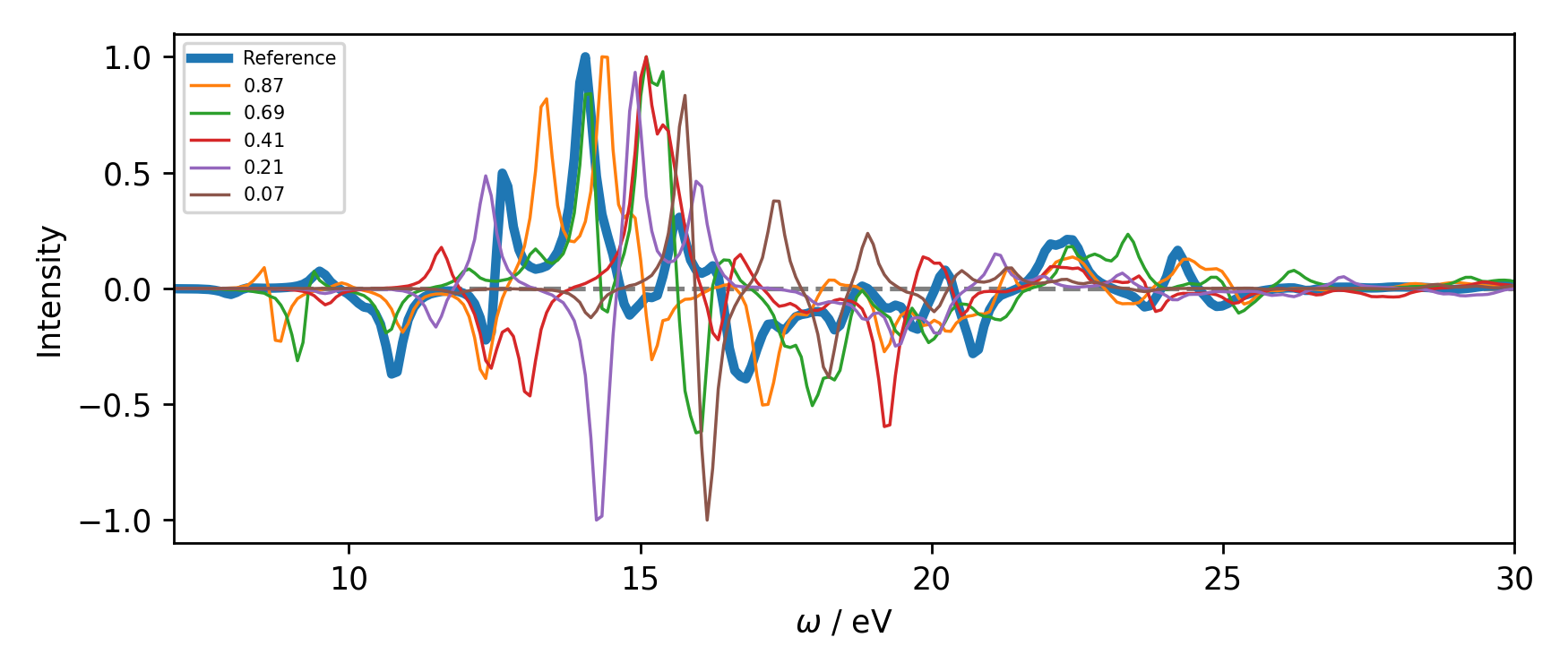}
    \caption{Reference and PNO ECD spectra of $(H_2)_4$ for five cutoffs: 
    [$1\times 10^{-10}$, $1\times 10^{-9}$, $1\times 10^{-8}$, $1\times 10^{-7}$, 
    $2\times 10^{-6}$] corresponding to $T_2$ ratios of 
    [$0.87$, $0.69$, $0.41$, $0.21$, $0.07$], respectively.}
    \label{fig:pno_ecd}
\end{figure}

Unlike with absorption, the ECD spectrum does not improve significantly when switching
to the PNO basis, as shown in Fig.~\ref{fig:pno_ecd}.
At a $T_2$ ratio of $0.87$, the overall \textit{shape} of the spectrum in the 10 eV to 
20 eV range more closely resembles that of the reference; however, the excitation
energies are, in some cases, even less accurate than those of smaller PNO spaces,
and the trend of lowering excited state energies with increased virtual space seen in 
Section~\ref{sss:abs} is no longer discernible. 

\begin{figure}
    \centering
    \includegraphics[scale=.8]{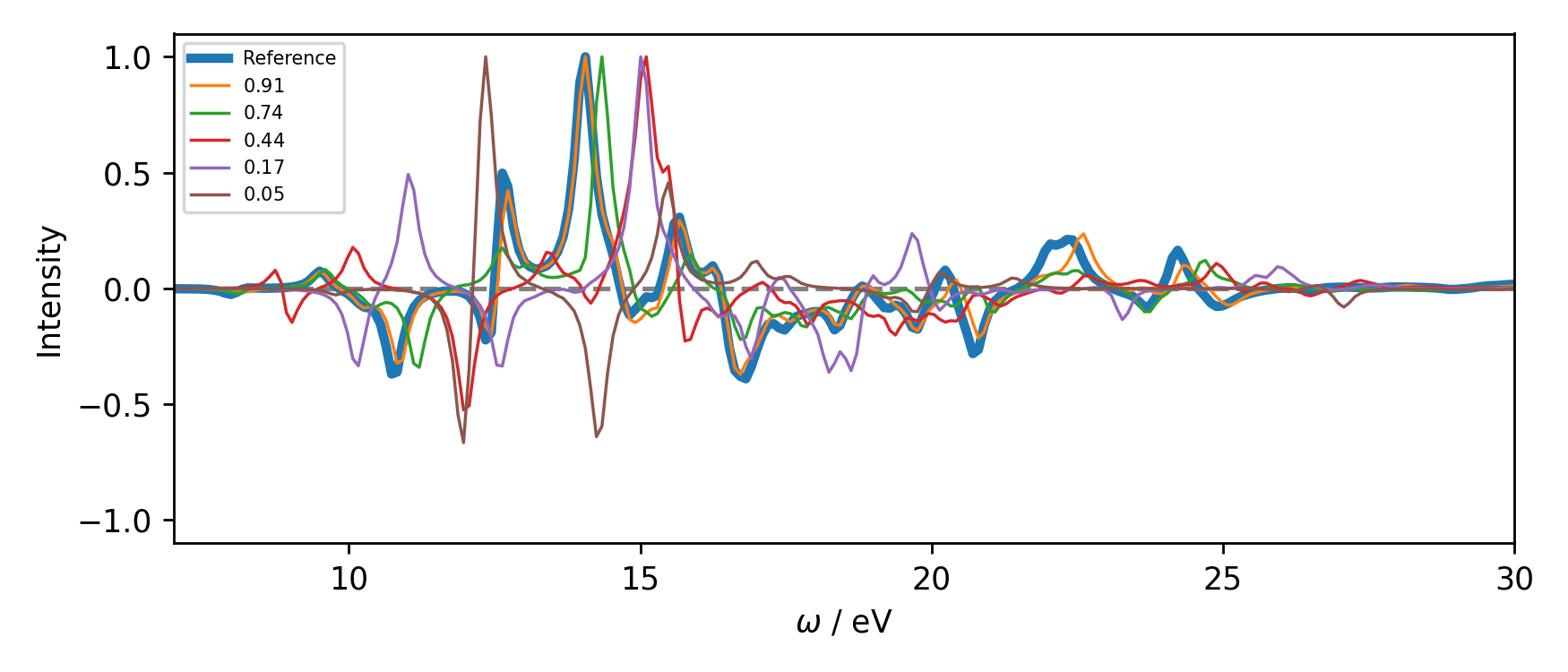}
    \caption{Reference and PNO++ ECD spectra of $(H_2)_4$ for five cutoffs: 
    [$1\times 10^{-9}$, $1\times 10^{-8}$, $1\times 10^{-7}$, 
    $1\times 10^{-6}$, $1\times 10^{-5}$] corresponding to $T_2$ ratios of 
    [$0.91$, $0.74$, $0.44$, $0.17$, $0.05$], respectively.}
    \label{fig:pnopp_ecd}
\end{figure}

Finally, Fig.~\ref{fig:pnopp_ecd} reports ECD spectra obtained upon truncation
of the PNO++ spaces used in the previous section. 

As expected based on frequency-domain
results,\cite{McAlexander2016,Kumar2017,Crawford2019,DCunha2021,DCunha2022}
ECD proves to be a greater challenge for the PNO++ method than the absorption.
However, there is still marked improvements compared to PAO and PNO. At a
$T_2$ ratio of 0.74, most distinguishing features of the spectrum below 20 eV
are present. The only notable exception to this is the Cotton feature
located near 12.5 eV, which is recovered when moving to a $T_2$ ratio of 0.91.
As with absorption, while one cannot truncate as aggressively as when
performing energy calculations, these results show a clear benefit to
including the response to the perturbation when building new virtual orbital
spaces for reduced-scaling techniques.

\subsection{Amplitude Dynamics} \label{ss:amps}
As evidenced by the preceding data, the truncated PAO and PNO virtual spaces do not
efficiently model the wave function in the presence of a perturbing EMF. As noted in
Section~\ref{sss:ecd}, these shortcomings are well-documented in the case of response
theory. Including the effects of the perturbation through the PNO++ approach improves
results, though possible truncations still fall short of those used for energy calculations. 
However, a real-time formalism offers the opportunity to analyze the wave 
function dynamics in great detail, perhaps shedding light on \textit{where} and 
\textit{how} the locally correlated wave functions are deficient. The following
section will scrutinize the $t_\mu$ and $\lambda_\mu$ amplitudes of 
Eqs.~(\ref{eq:t_mu}) and (\ref{eq:l_mu}), respectively, in order to determine the 
important fluctuations in the wave function and whether these spaces sufficiently 
capture these changes.

\begin{figure} 
    \centering
    \includegraphics[scale=.6]{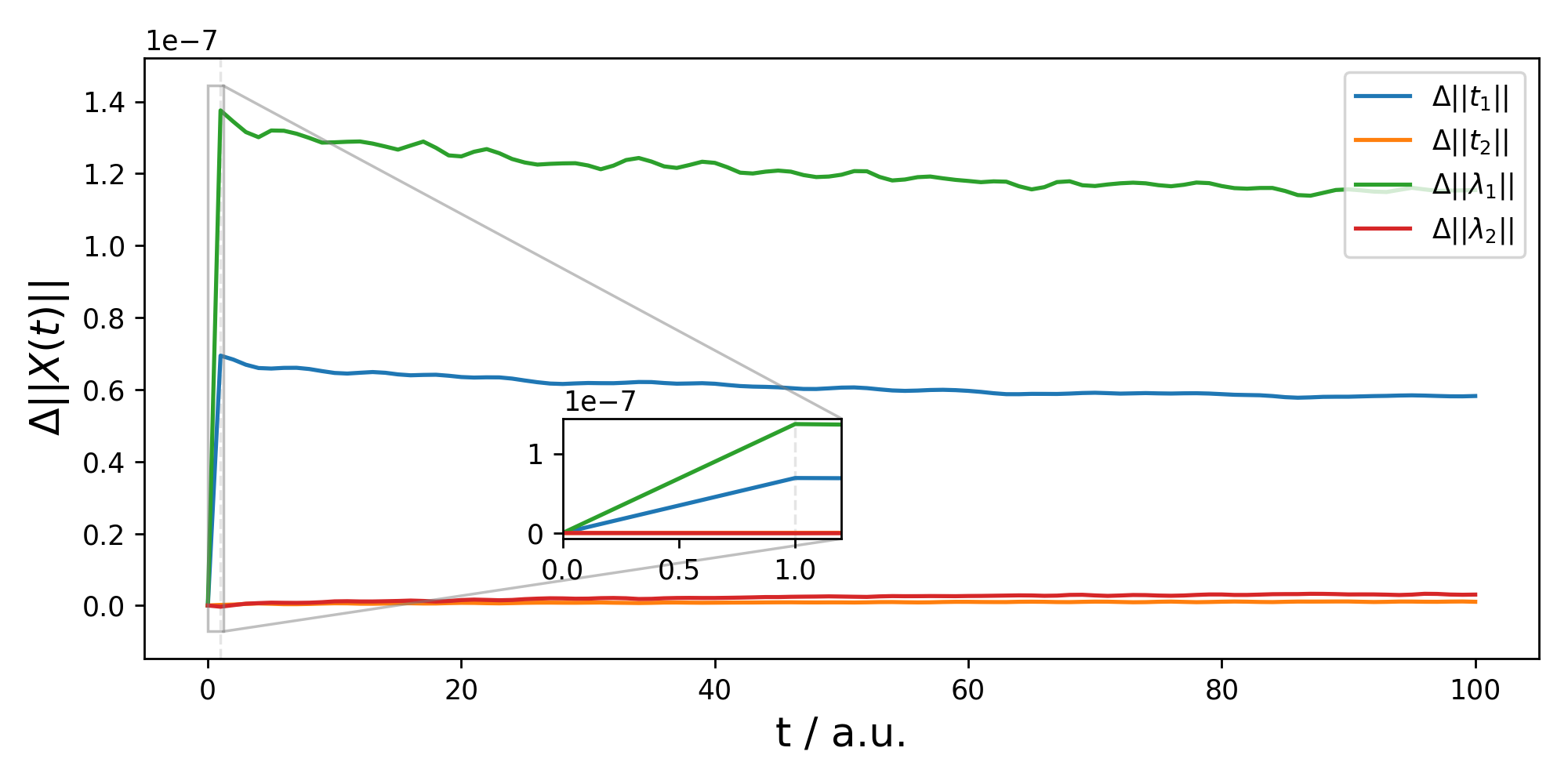}
    \caption{Time-dependent change in the norm of the amplitude 
    tensors relative to the ground-state amplitudes.
    Field and step parameters remain unchanged, and 
    the amplitude norm is taken at every 1 a.u.}
    \label{fig:norm}
\end{figure}

Response to external perturbations by the CC amplitudes gives rise to
dynamic energetics and properties. In the past, distributions of perturbed amplitudes
(relative to their ground-state counterparts) have been used to justify the 
difficulty in computing response functions with local correlation methods in the 
frequency domain. 
\cite{McAlexander2016,Crawford2019,DCunha2021} 
However, initial findings show that in RTCC, the relative distribution of amplitudes 
by magnitude is not significantly impacted.\cite{Crawford2019} 
Despite this, typical means of exploiting amplitude
sparsity have been shown to be inefficient by the preceding sections. 
First, to understand the response of 
the amplitudes to the external perturbation, we plot the 
change in the norm of the singles- and doubles-amplitude tensors relative to the unperturbed ground-state amplitudes 
as a function of time in Fig.~\ref{fig:norm}.
Results for the reduced-scaling spaces are identical to those
for the MO space, as the unitary transformations resulting from untruncated localized virtual spaces
in Eqs.~(\ref{eq:rotate}) preserves the tensor norm.

Fig.~\ref{fig:norm} shows that the largest component of the response by the wave function
is predominantly within the singles amplitudes $t_1$ and $\lambda_1$. This is 
consistent with the notion that singles are paramount for the computation of 
response properties.\cite{Christiansen1995,Koch1997} However, the density in Eq.~(\ref{eq:pair_D})
does not include any contributions by singles, due to being built from MP2-level
(perturbed) amplitudes where singles do not contribute until at least the second order 
in the wave function and fourth order in the energy. This suggests that even in schemes
that seek to include the EMF perturbation in the construction of the reduced virtual space,
such as PNO++, response of the singles should be considered, and this may provide significant
improvements.

\begin{figure}
    \begin{subfigure}{.5\textwidth}
        \centering
        \includegraphics[scale=0.5]{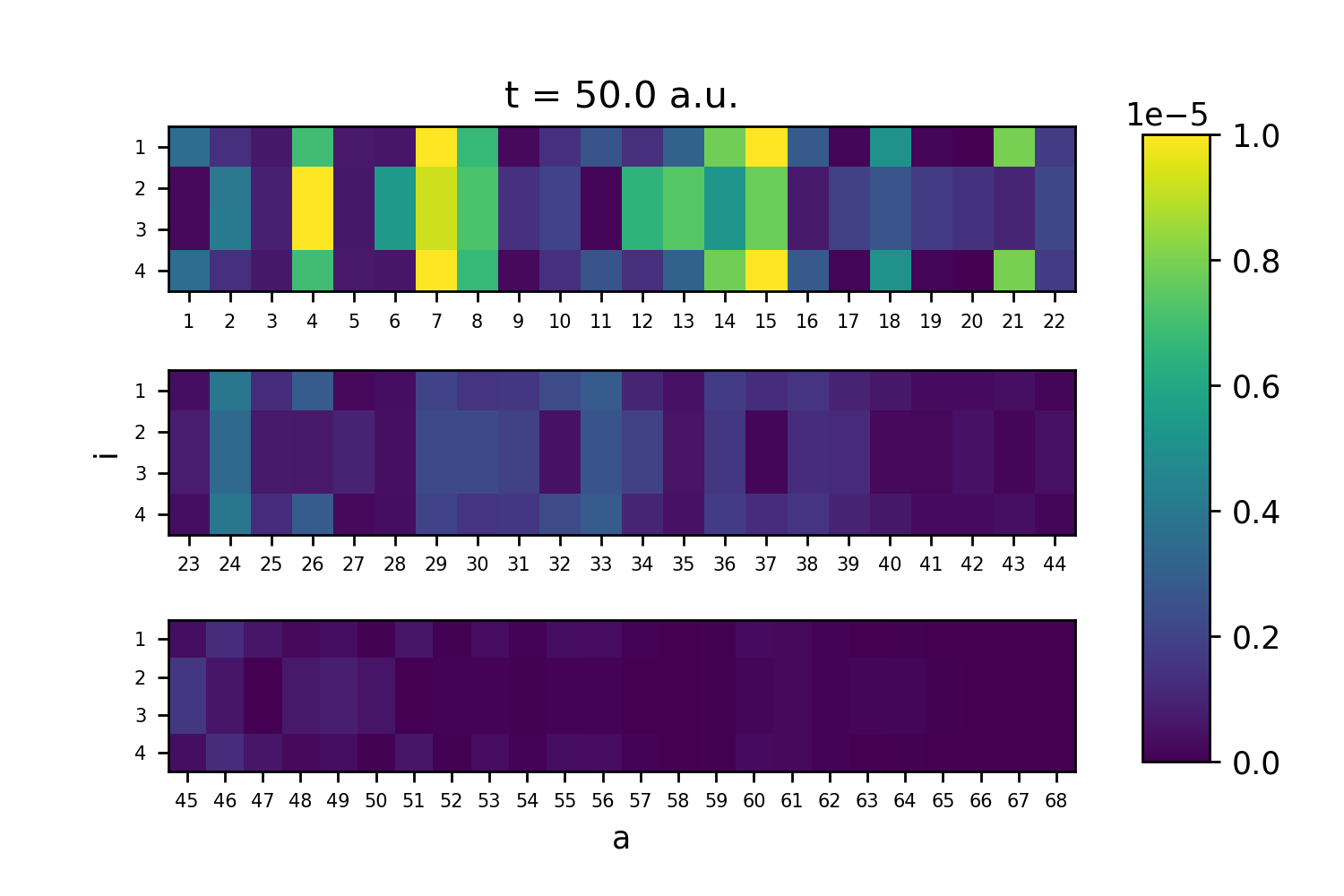}
        \caption{}
        \label{fig:MO_l1_50}
    \end{subfigure}%
    \begin{subfigure}{.5\textwidth}
        \centering
        \includegraphics[scale=0.5]{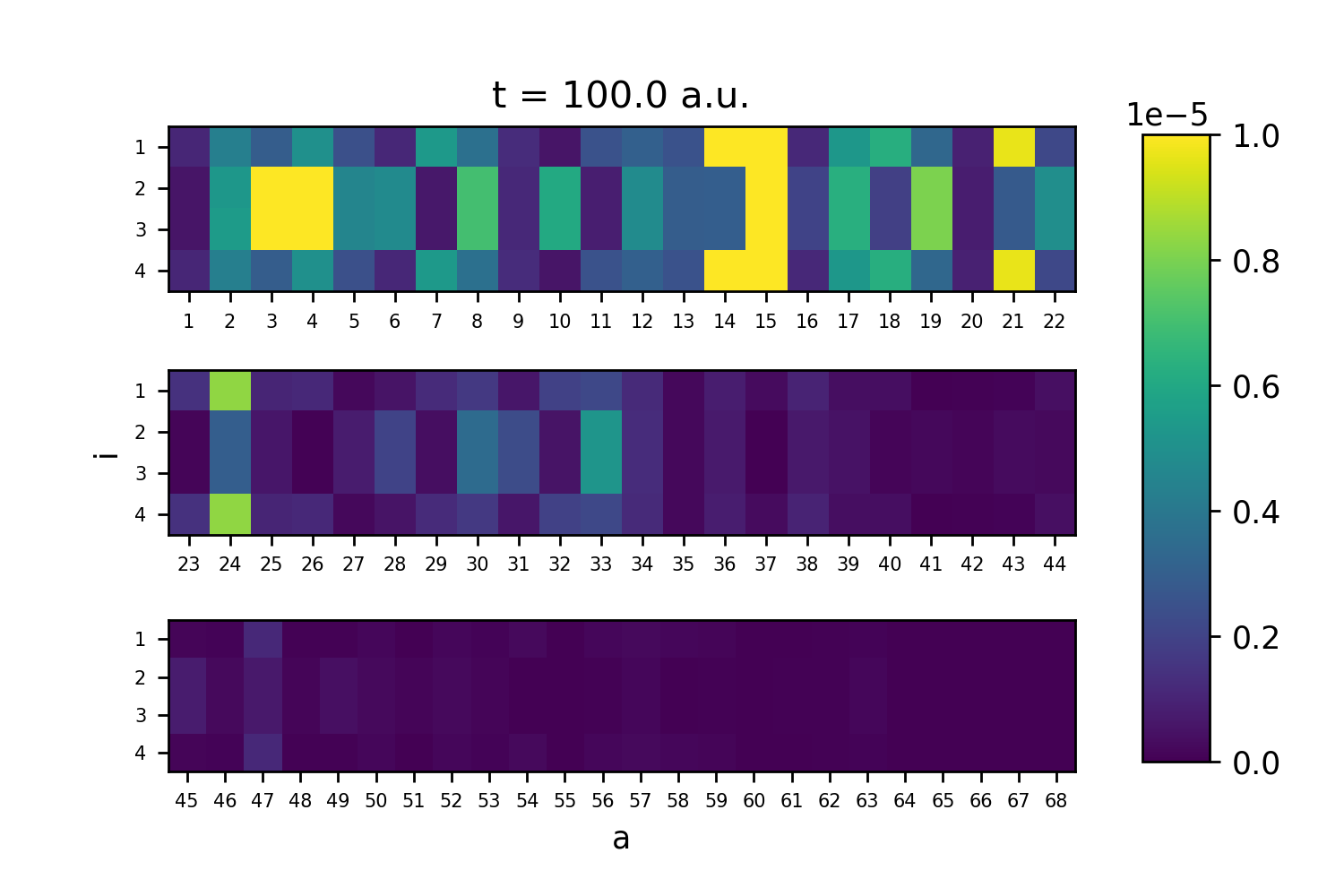}
        \caption{}
        \label{fig:MO_l1_100}
    \end{subfigure}
    \caption{MO-basis $\lambda_1$ amplitude deviations from $t = 0$ after 
    (a) 50 a.u. and (b) 100 a.u.
    of time propagation. Each row contains the same four localized occupied orbital indices
    and a subset of virtual indices as indicated by the x-axis labels, given in order of increasing orbital energy.}
    \label{fig:amps}
\end{figure}

In addition to the matrix norm, we can also inspect the individual amplitudes to track
their evolution in time. The heat maps in Fig.~\ref{fig:amps} show the difference 
in $\lambda_1$ amplitudes, relative to the ground state, 
for $t = 50 a.u.$ and $t = 100 a.u.$, to investigate deviations in the amplitudes
far from the initial electric field pulse.
The amplitudes are ordered by the energies of the associated MOs.  Those
amplitudes that experience significant oscillations vary throughout the
simulation, though there are several discernible trends. First, most large
amplitude deviations are associated with all occupied orbitals simultaneously.
This is due to the relatively small size of the system, with only four
occupied orbitals, all of which are likely important in the description of the
ground- and excited-state wave functions. Second, at any given time during
the propagation, a large number of amplitudes do not significantly deviate
from their ground-state values.  This supports the notion that the relative
sparsity of the wave function is maintained within the amplitudes throughout the simulation, but
this sparsity is distributed differently throughout the amplitude tensors as
the wave function is propagated.

A third trend is that amplitudes that respond strongly tend to be associated with 
low-energy virtual orbitals. Chemical intuition would suggest that energetically 
low-lying molecular orbitals will be the most involved in electronic excitations, particularly for valence states.
However, while amplitude responses are indeed larger for lower-energy virtual orbitals, 
smaller amplitude 
deviations in Fig.~\ref{fig:amps} extend far into the virtual space. This explains
the difficulty of simply truncating with respect to orbital energy: the 
high-energy MOs are still important to the time-evolution of the wave function 
in the presence of an EMF. 

\begin{figure}
    \begin{subfigure}{.5\textwidth}
        \centering
        \includegraphics[scale=0.5]{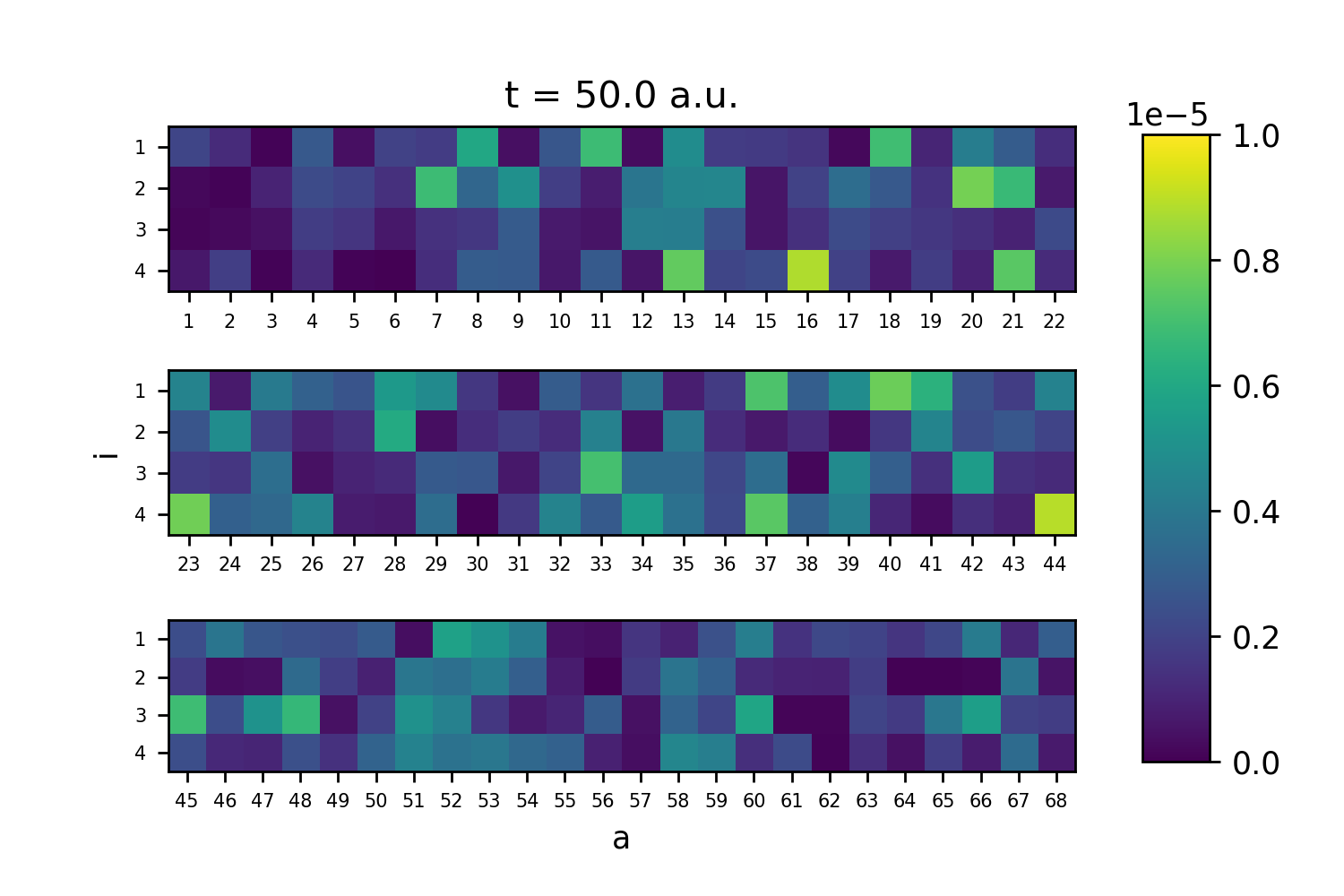}
        \caption{}
        \label{fig:PNO_l1_50}
    \end{subfigure}%
    \begin{subfigure}{.5\textwidth}
        \centering
        \includegraphics[scale=0.5]{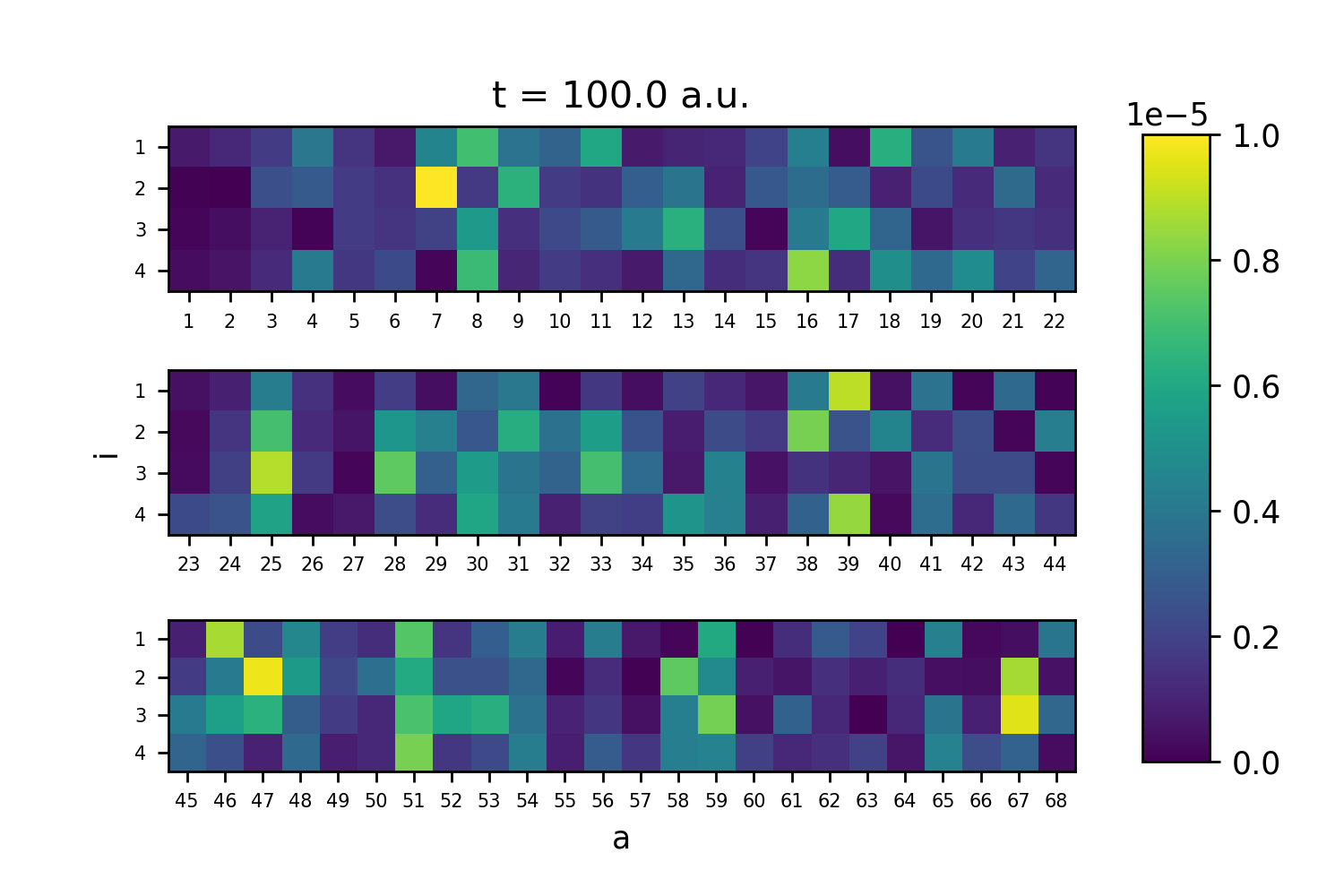}
        \caption{}
        \label{fig:PNO_t1_100}
    \end{subfigure}
    \caption{PNO-basis $\lambda_1$ amplitude deviations from $t = 0$ after 
    (a) 50 a.u. and (b) 100 a.u.
    of time propagation. Each row contains the same four occupied orbital indices
    and a subset of virtual indices as indicated by the x-axis labels, given in order of increasing orbital energy.}
    \label{fig:pno_amps}
\end{figure}

Fig.~\ref{fig:pno_amps} shows the $\lambda_1$ amplitudes for the same simulation,
rotated into the untruncated PNO basis using $Q_{ii}$ as defined in
Eq.~(\ref{eq:Q_pno}).
(It should be noted that, due to redundancy in the AO-based virtual 
spaces for each pair, PAO-basis amplitudes cannot be compared directly in 
this manner.)
It can be immediately seen that the amplitude deviations
are significantly less sparse in the PNO basis after the application of the EMF. 
Many more amplitudes exhibit
perceivable differences, and strong deviations (magnitudes approaching 
$1\times 10^{-5}$) are no longer present. This is a clear demonstration
of the issue with truncating orbital spaces based on the present criterion ---
rather than exploiting sparsity, the amplitude tensors have become less sparse.

\begin{figure}
    \begin{subfigure}{.5\textwidth}
        \centering
        \includegraphics[scale=0.5]{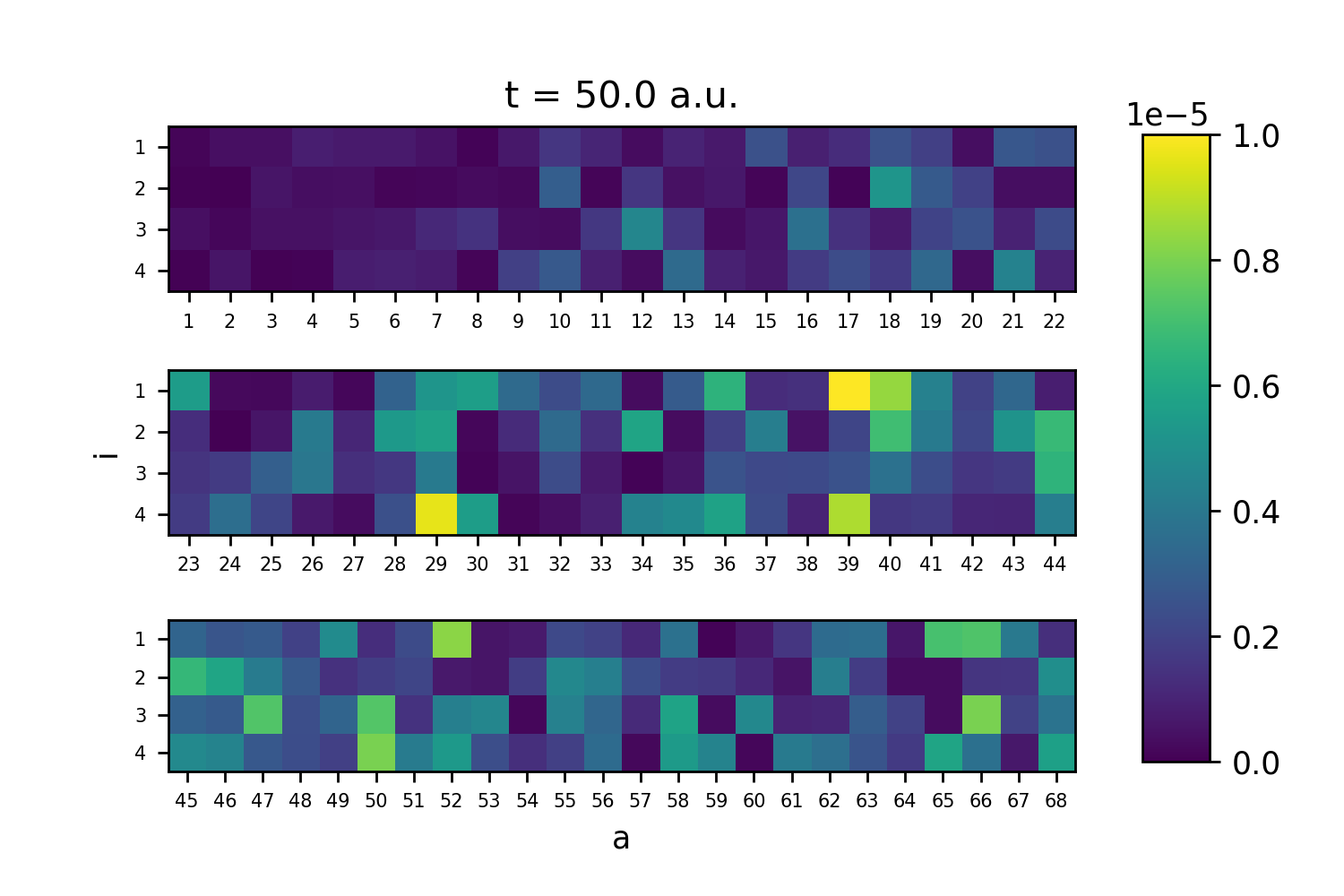}
        \caption{}
        \label{fig:PNOPP_l1_50}
    \end{subfigure}%
    \begin{subfigure}{.5\textwidth}
        \centering
        \includegraphics[scale=0.5]{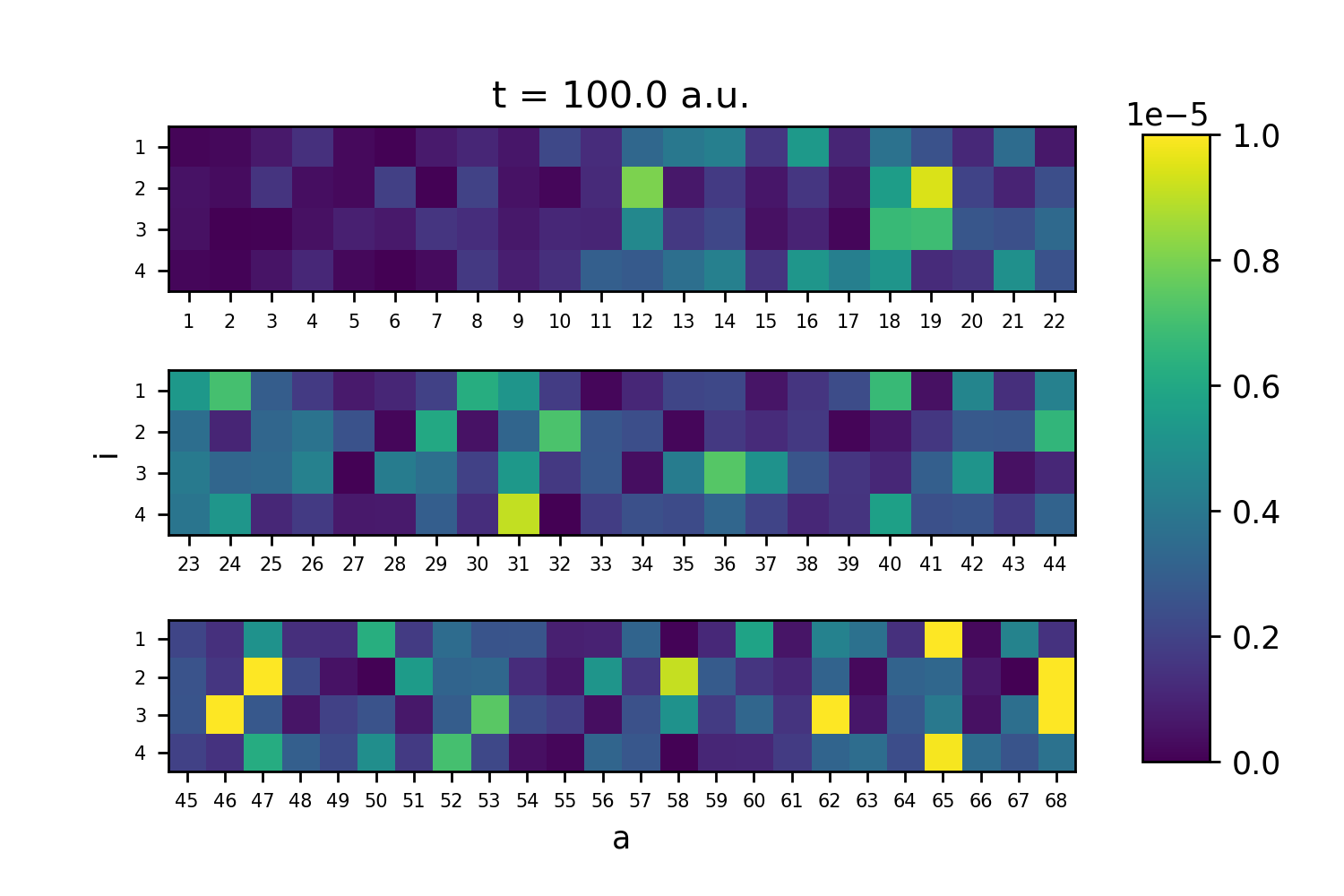}
        \caption{}
        \label{fig:PNOPP_t1_100}
    \end{subfigure}
    \caption{PNO++-basis $\lambda_1$ amplitude deviations from $t = 0$ after 
    (a) 50 a.u. and (b) 100 a.u.
    of time propagation. Each row contains the same four occupied orbital indices
    and a subset of virtual indices as indicated by the x-axis labels.}
    \label{fig:pnopp_amps}
\end{figure}

Finally, amplitude plots in the PNO++ basis are shown in Fig.~\ref{fig:pnopp_amps}.
We find that some sparsity has been retained; however, these amplitudes are
still not as sparse or as ordered as those in Fig.~\ref{fig:amps}. The
improvements of the PNO++ basis relative to the PNO basis are likely due to
this more efficient representation of the wave function amplitudes. 

\begin{figure}
    \centering
    \includegraphics[scale=0.75]{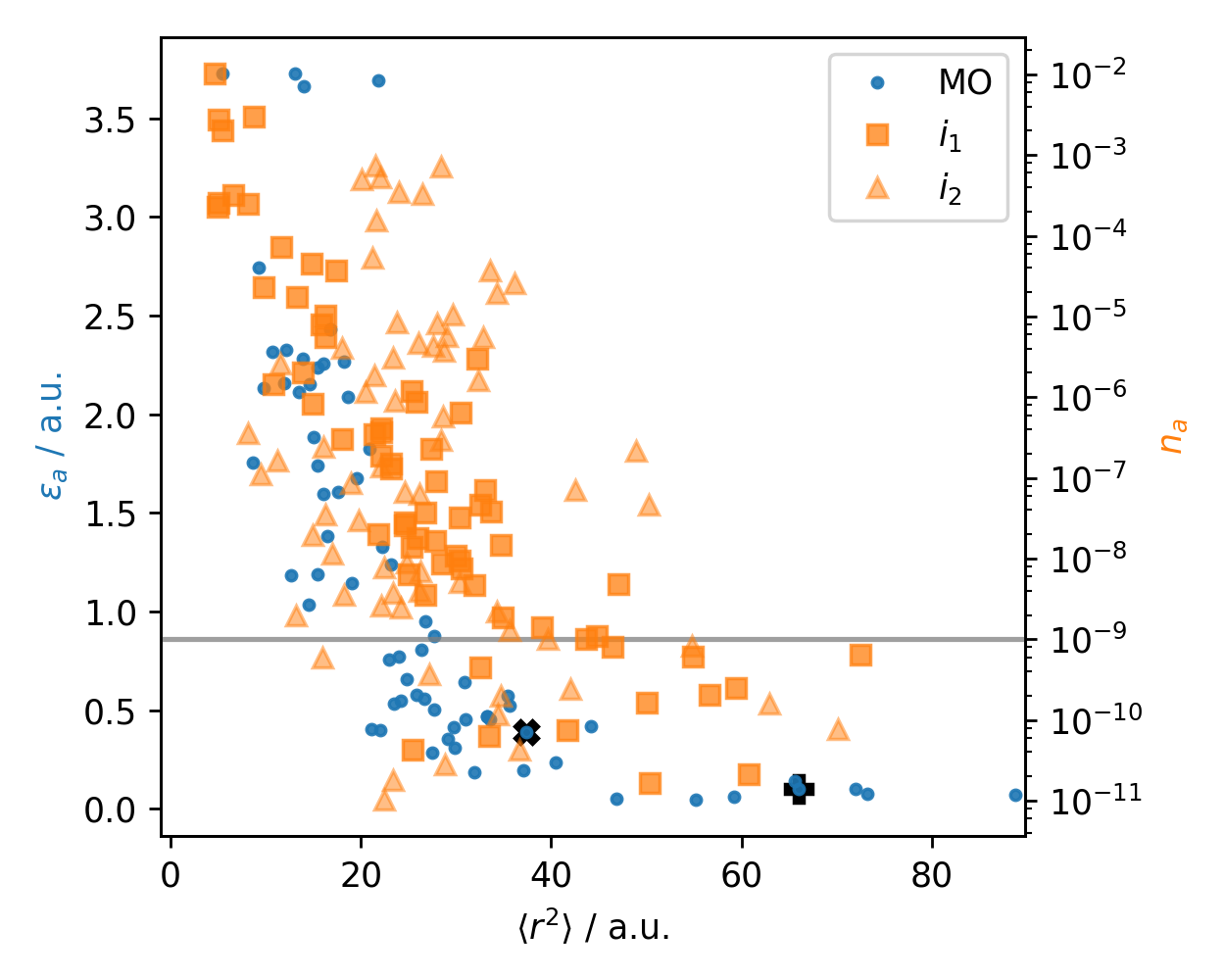}
    \caption{Virtual MO energy $\epsilon_a$ (left-hand axis) and PNO
    occupation number $n_a$ (right-hand axis, plotted on a log scale)
    for unique PNO spaces $i_1$ and $i_2$ 
    versus orbital extent.
    Virtual MOs 7 and 15 are denoted by a solid $\boldsymbol{+}$ 
    and $\boldsymbol{\times}$, respectively.
    The horizontal line denotes a PNO cutoff 
    of $1\times 10^{-9}$.}
    \label{fig:extent}
\end{figure}
The virtual orbital spatial extent has been used previously\cite{Kumar2017,DCunha2021} 
to estimate the ability of locally correlated spaces to describe the 
diffuse regions of electron density that are important for response
properties. 
Fig.~\ref{fig:extent} shows the virtual MO energy $\epsilon_a$ and the 
PNO occupation number $n_a$ plotted against the orbital extent
$\langle r^2 \rangle$ in a.u. 
In the PNO basis, a unique virtual space is prepared
for every occupied pair, resulting in 10 unique spaces for the four occupied
spatial orbitals $i$. 
However, for transforming
a single orbital index, we only require the diagonal rotation matrices,
\textit{i.e.}, $Q_{ii}$. There are four such spaces; however, by symmetry,
only two are unique, and both are included in Fig.~\ref{fig:extent}. 

Truncation of the PNO space begins from the bottom of Fig.~\ref{fig:extent}, 
e.g.\ at an occupation number cutoff of $1\times 10^{-9}$ (indicated by a horizontal
line), all orbitals below this line are neglected in the PNO space. 
This results in the PNO space with a $T_2$ ratio of 0.69.
From these data, it is clear
that even modest truncation of the virtual space neglects the diffuse 
regions of the wave function, which are important for excited-state
properties in systems with significantly delocalized characteristics,
such as systems containing Rydberg-type excitations.

\begin{figure}
    \centering
    \includegraphics[scale=0.75]{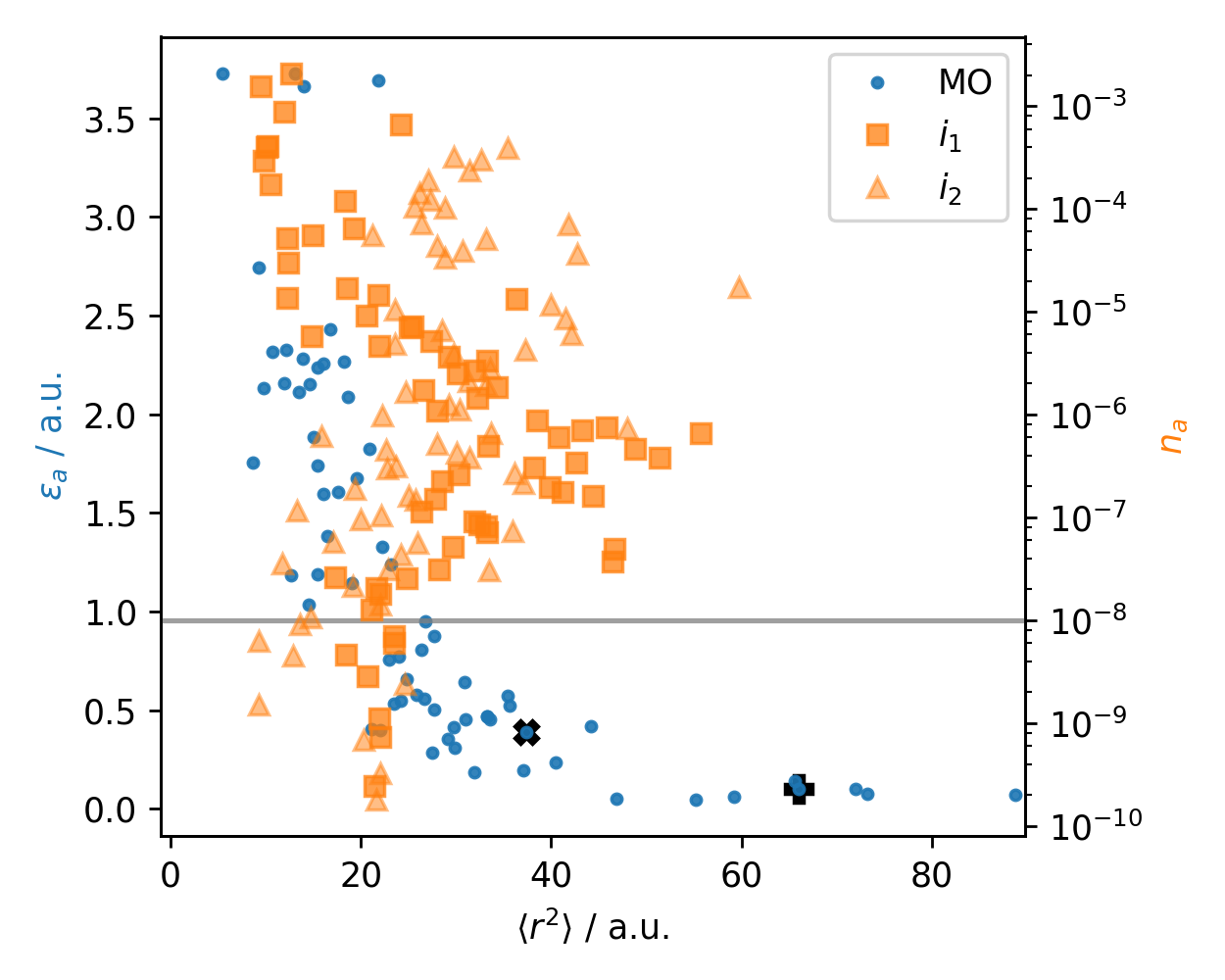}
    \caption{Virtual MO energy $\epsilon_a$ (left-hand axis) and PNO
    occupation number $n_a$ (right-hand axis, plotted on a log scale)
    for unique PNO++ spaces $i_1$ and $i_2$ 
    versus orbital extent.
    Virtual MOs 7 and 15 are denoted by a solid $\boldsymbol{+}$ 
    and $\boldsymbol{\times}$, respectively.
    The horizontal line denotes a PNO++ cutoff 
    of $1\times 10^{-8}$.}
    \label{fig:pnopp_extent}
\end{figure}

The same analysis can be performed for the PNO++ virtual spaces
in Fig.~\ref{fig:pnopp_extent}.
The horizontal line indicates a cutoff of $1\times 10^{-8}$, corresponding
to a $T_2$ ratio of 0.74. Below this threshold, at which point the quality
of the absorption and ECD spectra in Figs.~\ref{fig:pnopp_abs} and 
\ref{fig:pnopp_ecd} begins to deteriorate rapidly, more diffuse orbitals begin
to be neglected. This strongly alludes to the nature of the excited states,
and suggests spatial extent may be an important criterion to consider in future work.

Spatial extent alone may not be a suitable criterion for truncation --- 
this would have a negative impact on the accuracy of the 
correlation energy, which is inherently local in nature for molecules with isolated ground states.
Additionally, spatially compact orbitals may very well contribute 
significantly to the response of the wave function. 
Figs.~\ref{fig:extent} and \ref{fig:pnopp_extent} highlight virtual MOs 7 
($\boldsymbol{+}$) and 15 ($\boldsymbol{\times}$), which involve
the strongest deviations in 
Figs.~\ref{fig:MO_l1_50} and \ref{fig:MO_l1_100}, respectively.
These deviations represent a strong response of the wave function
to the perturbing field, implying they are of particular importance
when computing dynamic properties. 
Examination of the spatial extents of these orbitals in particular 
may shed light on the nature of the spatial distribution of 
orbitals necessary to describe the wave function response.
Virtual MO 7 appears at 66 a.u., while MO 15 is nearly half that at 37 a.u.
That these orbitals are of such varying extent
demonstrates that both diffuse and compact orbitals play a role in 
the wave function dynamics.


\section{Conclusions} \label{conc} We have presented the first application
of local correlation concepts to RTCC simulations. The popular PAO and PNO virtual
space localization schemes are applied to the calculation of dynamic electric
and magnetic dipole moments in the presence of an explicit electric field,
providing absorption and ECD spectra, respectively. 
For a helical H$_2$ tetramer test case, truncation of the
localized virtual space to successively larger fractions of the canonical
virtual space resulted in convergence to the canonical result; however, this
convergence is slow, and errors in excitation energies and intensity are
present even in some of the largest spaces tested, especially for ECD. 
The recently developed PNO++ scheme is shown to improve convergence significantly. This
corroborates the results of recent studies applying locally correlated
methods to the prediction of dynamic properties in the frequency domain
using response theory.  However, more work is needed to achieve the desired balance of accuracy and computational cost.

Examining the amplitude
dynamics during the propagation, it is shown that the $t_1$ and $\lambda_1$
amplitudes respond most strongly to the field -- a large increase in the
norm of these matrices is observed upon application of the field, followed
by a steady oscillation. The $t_2$ and $\lambda_2$ tensors, by comparison,
remain relatively static throughout. 
These oscillations are largely, but not completely, localized to a selection 
of only a few orbitals, as evidenced by consideration of 
time-dependent deviations in the $\lambda_1$ amplitudes from the ground-state.
In the localized virtual spaces tested,
these oscillations are delocalized throughout the $t_1$ and $\lambda_1$
matrices to different degrees, 
with the PNO++ virtual space retaining more sparsity than the PNO space.

Orbital extent alone cannot explain the shortcomings of 
the locally correlated methods explored here, though its effect is significant. 
These results provide an insight into the importance of 
singly-substituted determinants in the time-dependent wave function in the 
presence of an electric field, as well as a potential metric to gauge the 
performance of new localization schemes for frequency- or time-domain 
calculations of dynamic response properties.
In order to attain a balanced description of wave function components 
important for both energy and property calculations, the combination 
of appropriately determined spaces such as the combined PNO++ approach
has been fruitful.\cite{DCunha2021,DCunha2022} 
Still neglected in this approach
are the singles amplitudes, which are absent in the MP2 wave functions
used to approximate the occupied pair domains. Schemes to include 
these effects, such as approximate CC2-level $t_1$ guess amplitudes,
may further improve the space and allow greater flexibility for 
truncation.  In addition, rather than the broadband Dirac-delta pulse used for the external electric field, narrower
pulses that target specific frequency regions, e.g.\ for X-ray core excitations, could permit more aggressive truncation
of the wave function.
The prospect of utilizing these 
methodologies within the current framework is promising, and work is 
underway to explore their efficiency and efficacy.


\section{Supporting Information} \label{si}
An \texttt{.xyz} file containing the coordinates of the helical hydrogen
tetramer, (H$_2$)$_4$, is available (\texttt{h2\_4.xyz}).

\section{Acknowledgements} \label{ack}
This work was supported by supported by the U.S. National Science Foundation via
grant CHE-2136142. The authors gratefully acknowledge Advanced
Research Computing at Virginia Tech for providing computational resources
and technical support that have contributed to the results reported within
the paper.

\bibliography{lrtcc.bib}
\newpage
\begin{center}
{\large for Table of Contents use only}

\begin{figure}[h]
    \includegraphics[scale=0.7]{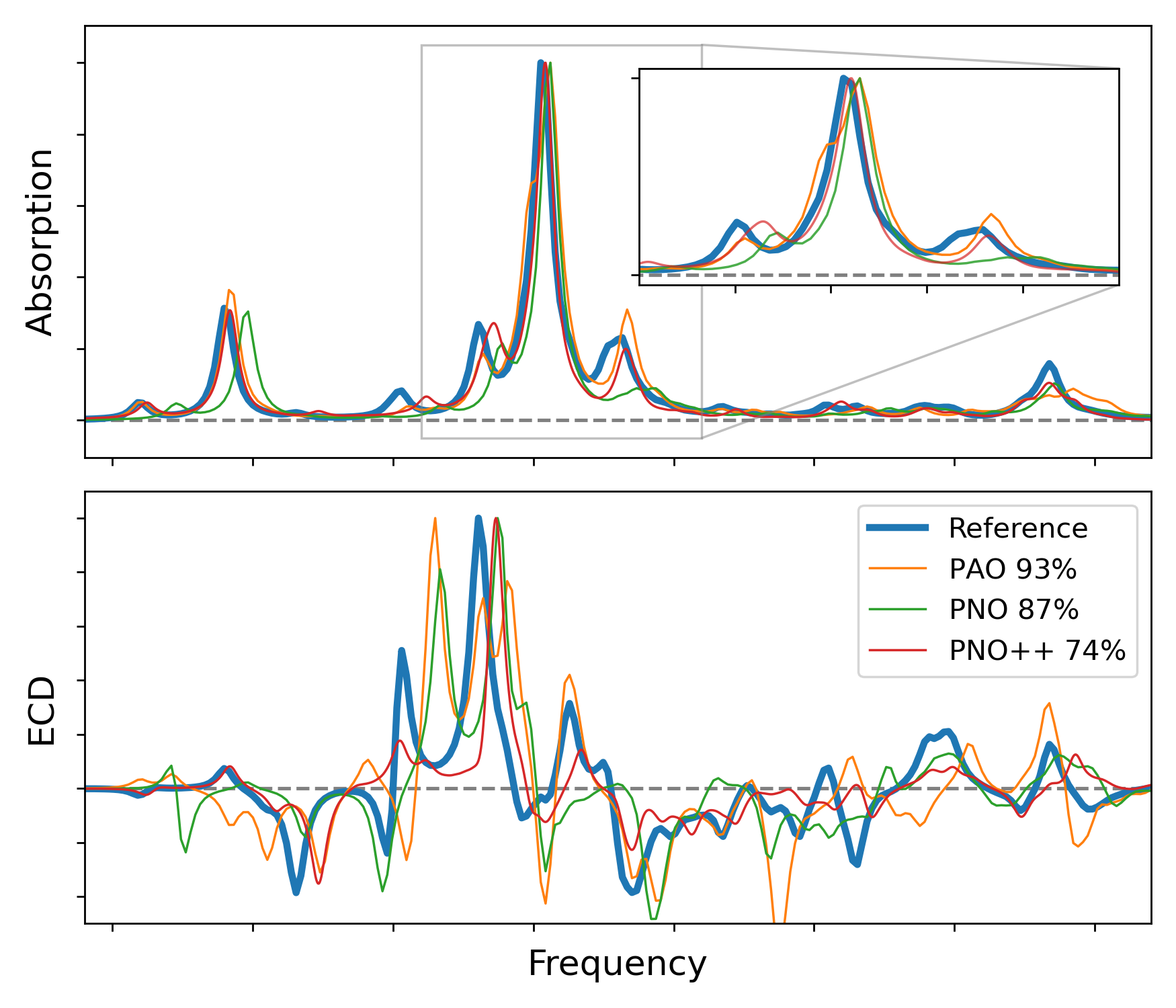}
    \caption{TOC Graphic}
\end{figure}
\vspace{2cm}
{\large Reduced Scaling Real-Time Coupled Cluster Theory}
\vspace{1.5cm}

{Benjamin G. Peyton, Zhe Wang, and T. Daniel Crawford}
\end{center}

\end{document}